\documentclass[12pt]{iopart}
\usepackage{iopams}
\usepackage{graphicx}
\usepackage{subfigure}
\newcommand{\beq}{\begin{equation}}
\newcommand{\eeq}{\end{equation}}
\newcommand{\bk}{{{\bf{k}}}}

\renewcommand{\br}{{{\bf{r}}}}

\newcommand{\bA}{{\bf{A}}}
\newcommand{\bB}{{\bf{B}}}
\newcommand{\bE}{{\bf{E}}}

\newcommand{\bq}{{\bf{q}}}
\newcommand{\bp}{{\bf{p}}}
\newcommand{\bb}{{\bf{b}}}

\newcommand{\beqa}{\begin{eqnarray}}
\newcommand{\eeqa}{\end{eqnarray}}
\newcommand{\pdg}{{\vphantom \dag}}
\newcommand{\dg}{{\dag}}
\renewcommand{\bnabla}{{\boldsymbol \nabla}} 
\renewcommand{\bsigma}{{\boldsymbol \sigma}}
\renewcommand{\btau}{{\boldsymbol \tau}}
\renewcommand{\bOmega}{{\boldsymbol \Omega}}
\renewcommand{\bGamma}{{\boldsymbol \Gamma}}
\newcommand{\upa}{\uparrow}
\newcommand{\da}{\downarrow} 

\newcommand{\cH}{{\cal H}}

\newcommand{\cD}{{\cal D}}
\newcommand{\eqref}{\ref}
\begin{document}
\topical{Chiral anomaly and transport in Weyl metals}
\author{A.A. Burkov}
\address{Department of Physics and Astronomy, University of Waterloo, Waterloo, Ontario 
N2L 3G1, Canada} 
\date{\today}
\begin{abstract}
We present an overview of our recent work on transport phenomena in Weyl metals, which 
may be connected to their nontrivial topological properties, particularly to chiral anomaly. 
We argue that there are two basic phenomena, which are related to chiral anomaly in Weyl metals: 
Anomalous Hall Effect (AHE) and Chiral Magnetic Effect (CME). While AHE is in principle present in any ferromagnetic 
metal, we demonstrate that a magnetic Weyl metal is distinguished from 
an ordinary ferromagnetic metal by the absence of the extrinsic and the Fermi surface part of the intrinsic 
contributions to the AHE, as long as the Fermi energy is sufficiently close to the Weyl nodes. The AHE in a Weyl metal is thus shown to be a purely 
intrinsic, universal property, fully determined by the location of the Weyl nodes in the first Brillouin zone. 
In other words, a ferromagnetic Weyl metal may be thought of as the only example of a ferromagnetic metal with a purely intrinsic AHE. 
We further develop a fully microscopic theory of diffusive magnetotransport in Weyl metals. 
We derive coupled diffusion equations for the 
total and axial (i.e. node-antisymmetric) charge densities and show that chiral anomaly manifests as 
a magnetic-field-induced coupling between them. We demonstrate that 
an experimentally-observable consequence of CME in magnetotransport 
in Weyl metals is a quadratic negative magnetoresistance, which will dominate all 
other contributions to magnetoresistance under certain conditions and may be regarded as a smoking-gun transport characteristic, unique to Weyl metals.
\end{abstract}
\maketitle
\section{Introduction}
\label{sec:1}
The last decade has seen an explosion of research in topologically-nontrivial states of matter, following the remarkable discovery of topological insulators (TI)~\cite{Kane05,Zhang06,Balents07,Kane07,Molenkamp07,Hasan09,Kane10,Qi10}. 
One normally associates topologically nontrivial properties with insulators, the spectral gap providing the rigidity and insensitivity to fluctuations, which are 
characteristic of these states of matter. 
One of the most recent notable developments in this field, however, has been the realization that even gapless metallic and semimetallic states 
may be topologically nontrivial in much the same sense as gapped insulators~\cite{Wan11,Ran11,Burkov11,Xu11}.
These recent developments have partly been anticipated in the earlier pioneering work of Volovik~\cite{Volovik}, promoting topological classification 
of all possible fermionic ground states, which shed new light on such wide-ranging phenomena as the robustness of the Fermi liquid and the hierarchy problem in particle physics. 

The recent work focussed on {\em Weyl semimetals}~\cite{Wan11,Ran11,Burkov11,Xu11}, which, in Volovik's classification, belong to the Fermi point 
universality class of fermionic vacua.
In the condensed matter context a Weyl point or node is a point of contact between two nondegenerate bands in the first Brillouin zone (BZ). Such electronic structure features have in 
fact been noticed and studied since the earliest days of the theory of solids~\cite{Herring,Abrikosov,Landau}, but only recently have their topological properties come into focus. 
Two important ingredients in the appearance of Weyl node features in the electronic structure are broken time reversal (TR) or inversion (I) symmetries and 
dimensionality of space. The broken symmetry requirement is a consequence of the well-known Kramers theorem. If both TR and I symmetries are present, the band eigenstates 
in a solid must be (at least) doubly degenerate at every value of the crystal momentum in the first BZ. 
Contact can then occur only between {\em pairs} of bands, which is in general impossible without fine-tuning.
The contact between two nondegenerate bands is, however, possible generically in three spatial dimensions (3D). 
Indeed, near the point of contact, the momentum-space Hamiltonian must have the following form, dictated by the so-called Atiyah-Bott-Shapiro 
construction~\cite{Horava}
\beq
\label{eq:1}
H = \pm v_F \bsigma \cdot \bk, 
\eeq
where the triplet of Pauli matrices $\bsigma$ describes the pair of touching bands, the sign in front refers to two possible {\em chiralities} of the band contact point, and we have subsumed any possible spatial anisotropies into the definition of the crystal momentum $\bk$. We will use $\hbar = c =1$ units throughout this paper, except in some of the final results. 
The point of band degeneracy occurs when all three components of the crystal momentum vanish. 
This is why three-dimensionality is important: we must have three momentum components available as ``tuning parameters" to create a point of degeneracy between the two bands. 
A naively analogous degeneracy point in 2D graphene in fact does not exist, as was pointed out in the seminal work of Kane and Mele~\cite{Kane05}, 
which started the field of TI. 
It only looks like a degeneracy point due the smallness of the spin-orbit (SO) interactions in graphene, which always open up a small gap and destroy the degeneracy. 

The presence of Weyl degeneracy nodes is, in principle, a common feature of the electronic structure of all 3D magnetic or noncentrosymmetric materials. 
In most cases, however, their presence is of no consequence, since they will generically occur far away from the Fermi energy~\cite{Doron13}. 
{\em Weyl semimetal} refers to a situation in which the Fermi level exactly coincides with the Weyl nodes and no other states are present at the Fermi energy. 
This situation is not generic and occurs only under special, but not entirely uncommon circumstances.

To identify these, it is convenient to start from the situation in which both TR and I symmetries are present.
TR symmetry requires that if a Weyl node is present at momentum $\bk$, another node with the same chirality must be present at $-\bk$. 
I symmetry, on the other hand, requires that if a Weyl node is present at $\bk$, a node of opposite chirality must be present at $-\bk$. 
This implies that in the presence of both TR and I, there may exist a pair of opposite-chirality Weyl nodes at the same TR and I-invariant 
crystal momentum. For concreteness, we will take this momentum to be at the BZ centre, i.e. at the $\Gamma$ point. 
The minimal $\bk \cdot \bp$ Hamiltonian, describing this situation, has the form
\beq
\label{eq:2}
H = v_F \tau^z \bsigma \cdot \bk + \Delta \tau^x, 
\eeq
where $\btau$ is an additional  set of Pauli matrices that describes the two nodes of opposite chiralities, i.e. $H$ has the form of the Hamiltonian 
of a 3D four-component Dirac fermion with ``mass" $\Delta$. 
The mass term $\Delta$ annihilates the two Weyl nodes, unless it vanishes by symmetry or is fine-tuned to zero.  
To realize the former situation, the states at the $\Gamma$ point must either form a four-dimensional irreducible representation of the space group of the crystal, 
as proposed to occur in $\beta$-cristobalite BiO$_2$~\cite{Kane12} and, with a few modifications (Dirac Hamiltonian is replaced by Luttinger Hamiltonian in this case), in some of the pyrochlore iridate materials~\cite{Krempa12,Balents13,Krempa14}. Or they must belong to two distinct two-dimensional representations, so that pairwise crossing, 
protected by crystal symmetry,
is possible, as realized in the recently 
discovered Dirac semimetal materials $\textrm{Na}_3 \textrm{Bi}$ and 
$\textrm{Cd}_2 \textrm{As}_3$~\cite{Fang12,Fang13,Cava13,Shen13,Hasan13}. 

The latter situation, i.e. with $\Delta$ fine-tuned to zero, occurs at the transition point between an inversion-symmetric 3D TI and a normal insulator (NI)~\cite{Murakami07,Burkov11,Ando11,Ando14}. In the present review we will focus exclusively on this case. 
Our results, however, do not depend on this choice and are applicable to all realizations of Dirac and Weyl semimetals. 
As mentioned above, Weyl semimetal may be obtained from Dirac semimetal by breaking either TR or I. We will focus on the TR-breaking case here, 
since it produces the simplest possible kind of Weyl semimetal, with only a single pair of Weyl nodes.  
Generalization to many pairs is in most cases straightforward, but it does complicate the technical details of course. 
Formally, this type of Weyl semimetal is obtained by adding a Zeeman spin-splitting term to the Dirac Hamiltonian Eq.~\eqref{eq:2}
\beq
\label{eq:3}
H = v_F \tau^z \bsigma \cdot \bk + \Delta \tau^x + b \sigma^z, 
\eeq
where the spin-splitting term may arise either from an external magnetic field, or from an intrinsic magnetic ordering, and we have chosen the 
$z$-direction to be the magnetization direction. 
To analyze this Hamiltonian, it is convenient to perform the following canonical (i.e. commutation relation preseving) transformation of the $\bsigma$ and $\btau$ operators
\beq
\label{eq:4}
\sigma^{\pm} \rightarrow \tau^z \sigma^{\pm},\,\, \tau^{\pm} \rightarrow \sigma^z \tau^{\pm}. 
\eeq
The Hamiltonian Eq.~\eqref{eq:3} then transforms to
\beq
\label{eq:5}
H = v_F (\sigma^x k_x + \sigma^y k_y) + (b + v_F k_z \tau^z + \Delta \tau^x) \sigma^z, 
\eeq
Diagonalizing the $v_F k_z \tau^z + \Delta \tau^x$ block of the Hamiltonian we obtain
\beq
\label{eq:6}
H_{\pm} = v_F (\sigma^x k_x + \sigma^y k_y) + m_{\pm}(k_z) \sigma^z, 
\eeq
where 
\beq
\label{eq:7}
m_{\pm}(k_z) = b \pm \sqrt{v_F^2 k_z^2 + \Delta^2}. 
\eeq
Eq.~\eqref{eq:6} looks like the Hamiltonian of a pair of 2D Dirac fermions with ``masses" $m_{\pm}(k_z)$, which depend 
on the $z$-component of the crystal momentum as a parameter. 
The mass $m_+(k_z)$ is always positive, but $m_-(k_z)$ may vanish and change sign if $b > \Delta$. 
If this is the case, $m_-(k_z)$ vanishes at $k_z^{\pm} = \pm k_0$, where 
\beq
\label{eq:8}
k_0 = \frac{1}{v_F} \sqrt{b^2 - \Delta^2}. 
\eeq
These points along the $z$-axis in momentum space, where the ``mass" $m_-(k_z)$ vanishes, are the locations of the Weyl nodes. 
The two bands, which are the eigenstates of the $H_-$ block of the Hamiltonian, are degenerate at those points.  
Since massive 2D Dirac fermions are associated with half-quantized Hall conductivity~\cite{Fisher94}
\beq
\label{eq:9}
\sigma_{xy}^{2D} = \frac{e^2}{2 h} \textrm{sign}(m), 
\eeq
it follows that the Weyl semimetal at $b > \Delta$ has a Hall conductivity given by
\beq
\label{eq:10}
\sigma_{xy} = \frac{e^2}{h} 2 k_0, 
\eeq
i.e. is equal to the distance between the Weyl nodes in units of $e^2/h$~\cite{Klinkhamer}.

It is easy to show that this Hall conductivity may be associated with chiral edge states. 
Indeed, suppose the sample is finite in the $y$-direction and we will take the sample to
occupy the $y < 0$ half-space. 
Replacing $k_y \rightarrow -i \partial/ \partial y$, the $H_-$ block of the Hamiltonian becomes
\beq
\label{eq:11}
H_- = - i v_F \frac{\partial}{\partial y} \sigma^y + v_F \sigma^x k_x + m_-(k_z,y) \sigma^z. 
\eeq
Let us first set $k_x = 0$ and look for a zero-energy solution of the Schr\"odinger equation (SE)
\beq
\label{eq:12}
H_- \Psi(k_z,y) = 0, 
\eeq
localized at the sample boundary $y = 0$. 
We look for a solution in the form 
\beq
\label{eq:13}
\Psi(k_z,y) = i \sigma^y e^{f(k_z,y)} \phi, 
\eeq
where $f(k_z,y)$ is a scalar function, while $\phi$ is a two-component spinor. 
Plugging this into the SE Eq.~\eqref{eq:12}, we obtain
\beq
\label{eq:14}
\left[v_F \frac{\partial  f(k_z,y)}{\partial y} + m_-(k_z,y) \sigma^x \right] \phi = 0. 
\eeq
The solution of Eq.~\eqref{eq:14}, satisfying our requirements is 
\beq
\label{eq:15}
f(k_z,y) = \frac{1}{v_F} \int_0^y m_-(k_z, y') d y', 
\eeq
and $\phi = | \sigma^x = -1 \rangle$, i.e. the eigenstate of $\sigma^x$ with eigenvalue $-1$. 
Thus we finally obtain the following result for the solution of Eq.~\eqref{eq:14}, localized at the 
sample boundary
\beq
\label{eq:16}
\Psi(k_z,y) = e^{\frac{1}{v_F} \int_0^y m_-(k_z,y') dy'} | \sigma^x = 1 \rangle. 
\eeq
This solution exists as a localized edge state as long as $m_-(k_z, y \rightarrow - \infty) \geq 0$, i.e. 
as long as $-k_0 \leq k_z \leq k_0$. 
For this reason it is called a Fermi arc~\cite{Wan11}. 
It is straightforward to see that the dispersion of this state in the $x$-direction is given by $\epsilon = v_F k_x$, i.e.
it is chiral. 

At this point we will wrap up the introductory part of this review and move on to a more detailed account of the transport 
properties of Weyl semimetals and metals, based on a more realistic microscopic model of Weyl semimetal. 
The rest of the paper is organized as follows. 
In Section~\ref{sec:2} we introduce a realistic microscopic model of a Weyl semimetal, based on a magnetically-doped
TI-NI multilayer heterostructure. 
In Section~\ref{sec:3} we discuss basic electromagnetic response properties of this model Weyl semimetal and introduce 
two distinct components of the response, which are directly related to the nontrivial topology of Weyl nodes: the Anomalous Hall Effect (AHE)~\cite{AHE1,AHE2} 
and the Chiral Magnetic Effect (CME). 
In Section~\ref{sec:4} we extend the theory of electromagnetic response of Weyl metals to diffusive transport regime. 
We demonstrate that magnetic Weyl metals are distinguished from ordinary ferromagnetic metals by a lack of impurity-scattering 
and Fermi-surface contributions to their anomalous Hall conductivity. Instead their anomalous Hall conductivity is determined 
by the distance between the Weyl nodes in momentum space and nothing else. 
We also show that CME manifests in the diffusive regime as a new kind of weak-field magnetoresistance, which is 
negative and quadratic in the field. This novel magnetoresistance is expected to occur in all types of Weyl metals
and may be regarded as their smoking-gun transport characteristic. 
We conclude in Section~\ref{sec:5} with a brief overview of the main results.

\section{Heterostructure model of Weyl semimetal}
\label{sec:2}
\begin{figure}[t]
\begin{center}
\includegraphics[width=15cm]{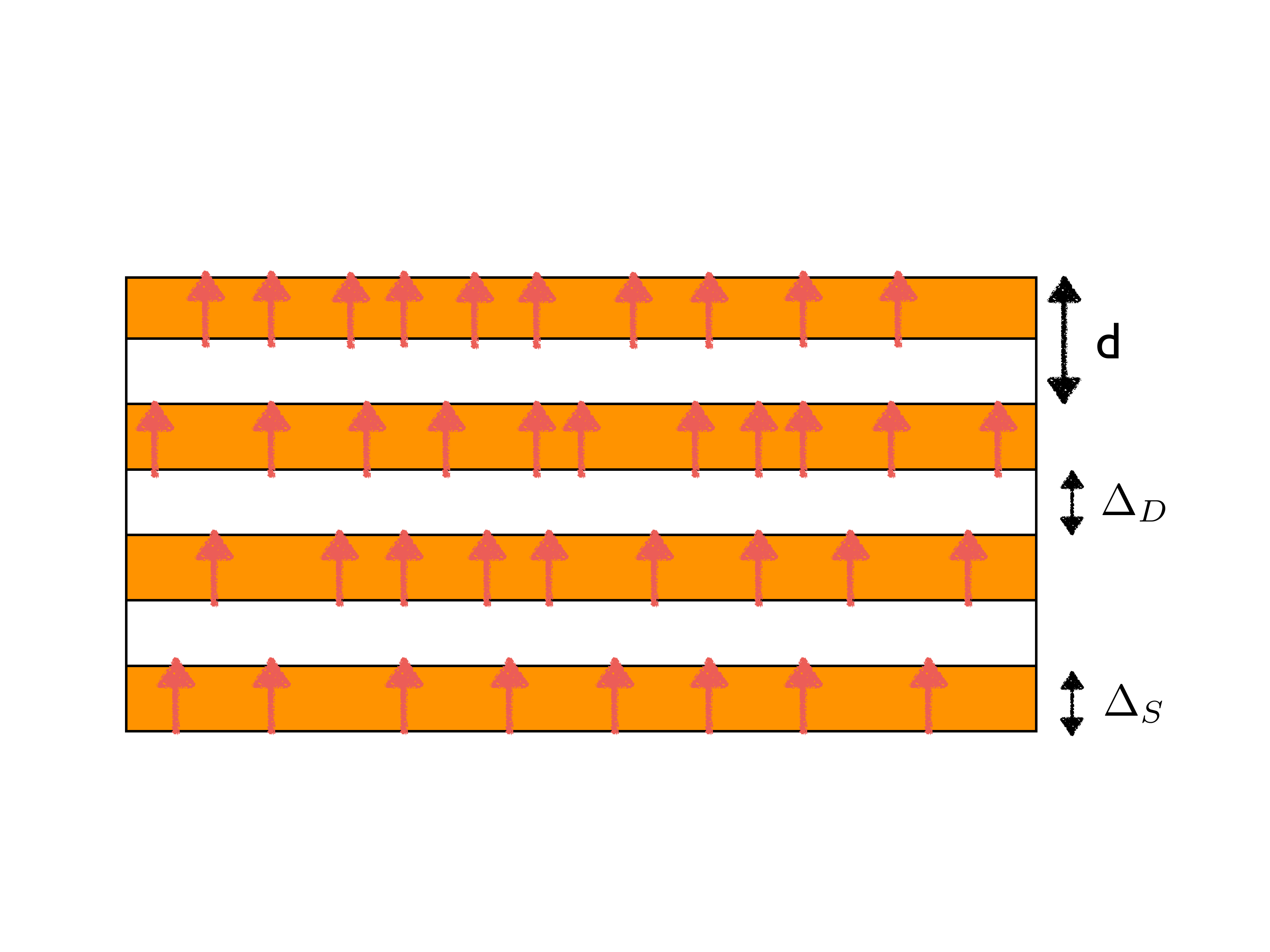}
\end{center}
\vspace{-2cm}
\caption{Cartoon of the heterostructure model of Weyl semimetal. Magnetized impurities are shown by arrows.}
\label{fig:1}
\end{figure}

As discussed in the Introduction, perhaps the most straightforward (at least theoretically) way to realize a Weyl semimetal phase is to 
break TR symmetry in a material, which in its nonmagnetic state is naturally poised near the phase transition between a TI and NI. 
One way to achieve this is to engineer a composite material, made of alternating thin layers of TI and NI, as shown in Fig.~\ref{fig:1}. 
This system may be regarded as a ``hydrogen atom" of Weyl semimetals: the most elementary yet realistic system, the description of which is simple enough that purely analytical 
theory of many phenomena is possible, as will be seen below.  

This system may be described as a chain of 2D Dirac surface states of the TI layers, which are coupled by tunnelling matrix elements 
$\Delta_S$ for a pair of surface states, belonging to the same TI layer, and $\Delta_D$ for a pair belonging to nearest-neighbor TI layers. 
We will assume, for concreteness, that $\Delta_{S,D} \geq 0$. 
The corresponding Hamiltonian has the form
\beq
\label{eq:17}
H = v_F \tau^z (\hat z \times \bsigma) \cdot \bk_{\perp} \delta_{i,j} + \Delta_S \tau^x \delta_{i,j} + \frac{1}{2}\ \Delta_D \left(\tau^+ \delta_{j, i+1} + \tau^- \delta_{j, i-1} \right), 
\eeq
where $i,j$ label the unit cells of the superlattice in the growth ($z$) direction, $\bk_{\perp} = (k_x, k_y)$ are the crystal momentum components, 
transverse to the growth direction, Pauli matrices $\btau$ act on the {\em which surface} pseudospin degree of freedom, while $\bsigma$ act in the spin degree of freedom. 
Diagonalizing the growth direction hopping part of the Hamiltonian by Fourier transform and performing the canonical transformation of Eq.~\eqref{eq:4}, we obtain 
\beq
\label{eq:18}
H = v_F (\hat z \times \bsigma) \cdot \bk + \hat \Delta(k_z) \sigma^z, 
\eeq
where 
\beq
\label{eq:19}
\hat \Delta(k_z) = \Delta_S \tau^x + \frac{\Delta_D}{2} \left(\tau^+ e^{i k_z d} + \textrm{h.c.} \right), 
\eeq
$d$ being the period of the superlattice in the growth direction.
$\hat \Delta(k_z)$ may now be diagonalized separately, as it commutes with the Hamiltonian. 
Its eigenvalues are given by $\pm \Delta(k_z)$, where 
\beq
\label{eq:20}
\Delta(k_z) = \sqrt{\Delta_S^2 + \Delta_D^2 + 2 \Delta_S \Delta_D \cos(k_z d)}. 
\eeq
The corresponding two-component spinor wavefunctions are
\beq
\label{eq:21}
|u^t(k_z) \rangle = \frac{1}{\sqrt{2}} \left(1, t \frac{\Delta_S + \Delta_D e^{-i k_z d}}{\Delta(k_z)} \right), 
\eeq
where $t = \pm$ labels the two eigenvalues. 
The spin block of the Hamiltonian may now be diagonalized as well, giving rise to the following eigenvalues
\beq
\label{eq:22}
\epsilon_{s t}(\bk) = s \epsilon_t(\bk) = s \sqrt{v_F^2(k_x^2 + k_y^2) + m^2_t(k_z)}, 
\eeq
where $s = \pm$ and 
\beq
\label{eq:23}
m_t(k_z) = t \Delta(k_z). 
\eeq
The double degeneracy of the band eigenvalues at every $\bk$ in Eq.~\eqref{eq:22} is the Kramers degeneracy, arising due to the 
presence of both TR and I symmetries. 
The corresponding eigenvectors are given by
\beq
\label{eq:24}
|v^{s t}(\bk) \rangle = \frac{1}{\sqrt{2}} \left(\sqrt{1 + s \frac{m_t(k_z)}{\epsilon_t(\bk)}}, - i s e^{i \phi} \sqrt{1 - s \frac{m_t(k_z)}{\epsilon_t(\bk)}} \right), 
\eeq
where $e^{i \phi} = \frac{k_x + i k_y}{\sqrt{k_x^2 + k_y^2}}$. 
The full four-component eigenvectors of the Hamiltonian Eq.~\eqref{eq:18} may be represented as a tensor product $|u^t(k_z)  \rangle$ and 
$|v^{s t}(\bk) \rangle$:
\beq
\label{eq:25}
|z^{s t}(\bk) \rangle = |v^{s t}(\bk) \rangle \otimes |u^t(k_z) \rangle. 
\eeq

The heterostructure, described by Eq.~\eqref{eq:17}, can exist in two distinct phases: strong 3D TI and an NI.  
This can be easily seen by computing the $Z_2$ index, using the method of Fu and Kane~\cite{Fu07}.  
Namely, we compute the eigenvalues of the parity operator $\tau^x$ (or $\tau^x \sigma^z$ after the canonical transformation of Eq.~\eqref{eq:4}) 
at two TR-invariant momenta in the first BZ: $\bGamma_1=(0,0,0)$ and $\bGamma_2=(0,0,\pi/d)$. 
Taking the product of the parity eigenvalues over all the occupied bands and over $\bGamma_{1,2}$, the $Z_2$ index is found to be
\beq
\label{eq:26}
(-1)^{\nu} = \textrm{sign}(\Delta_S - \Delta_D). 
\eeq
Thus the multilayer is a strong 3D TI when $\Delta_D > \Delta_S$ and an NI otherwise. 
The point $\Delta_S = \Delta_D$ marks the transition point between the 3D TI and NI. 
At this point, the gap at $\bGamma_2$ closes and the heterostructure is a Dirac semimetal. 

To obtain a Weyl semimetal, we break TR symmetry in the heterostructure, assuming that it is 
tuned close to the TI-NI transition point and the band gap $|\Delta_S - \Delta_D|$ is not large. 
The TR breaking may be accomplished by doping the heterostructure with magnetic impurities. 
At sufficient concentration, the impurities will form a ferromagnetic (FM) state, and we will assume that 
the magnetization points along the growth direction of the heterostructure (distinct, non-Weyl nodal states are 
obtained if the magnetization is in the $xy$-plane~\cite{Burkov11-2}). 
We will describe the FM state of the heterostructure by adding a term $b \sigma^z \delta_{i,j}$ to the Hamiltonian Eq.~\eqref{eq:17}. 
This splits the degeneracy of the $t = \pm$ Kramers doublet, modifying Eq.~\eqref{eq:23} as
\beq
\label{eq:27}
m_t(k_z) = b + t \Delta(k_z). 
\eeq
Taking the spin splitting $b$ to be always positive, the ``mass" $m_+(k_z)$ is then always positive, while $m_-(k_z)$ may vanish 
and change sign as a function of $k_z$, if the spin splitting is sufficiently large.  
This happens when $b \geq b_{c1} = |\Delta_S - \Delta_D|$. 
The Weyl node locations along the $k_x = k_y = 0$ line in the crystal momentum space are given by the solutions of the equation
\beq
\label{eq:28}
m_-(k_z) = b - \Delta(k_z) = 0, 
\eeq
which gives $k_{z}^{\pm} = \pi/d \pm k_0$, where 
\beq
\label{eq:29}
k_0 = \frac{1}{d} \arccos\left(\frac{\Delta_S^2 + \Delta_D^2 - b^2}{2 \Delta_S \Delta_D} \right). 
\eeq
When the spin splitting reaches the upper critical value $b_{c2} = \Delta_S + \Delta_D$, $k_0 = \pi/d$ and the 
Weyl nodes are annihilated again at the BZ edge. The resulting state is a 3D quantum anomalous Hall insulator 
with a quantized Hall conductivity~\cite{Halperin92}
\beq
\label{eq:30} 
\sigma_{xy} = \frac{e^2}{h d}. 
\eeq

\section{Electromagnetic response of Weyl semimetals}
\label{sec:3}
\subsection{General remarks}
\label{sec:3.1}
Topologically-nontrivial states of matter, such as 3D TI, and Weyl semimetals, have a distinguishing spectroscopic 
feature: the presence of metallic edge states. 
These surface metallic states are unusual and may only exist on the surface of a bulk 3D topological phase. 
As discussed in the Introduction, in the Weyl semimetal state, of interest to us here, the Fermi surface of this metal is not a closed 2D curve, as it must 
be in any regular 2D metal, but instead forms an open Fermi arc, with the end points at the locations of the bulk Weyl nodes, 
projected onto the surface BZ~\cite{Wan11,Potter14}. 

However, apart from such spectroscopic distinguishing features, topological phases often have unusual electromagnetic response characteristics, 
which are always of particular interest, as these are the manifestations of the unique quantum physics of these phases on macroscopic scales. 
In 3D TI, this unusual electromagnetic response may be described by the so-called $\theta$-term in the action of the electromagnetic field~\cite{Qi10}
\beq
\label{eq:31}
S_{\theta} = \frac{e^2 \theta}{32 \pi^2} \int d t \,\, d^3 r \epsilon^{\mu \nu \alpha \beta} F_{\mu \nu} F_{\alpha \beta} = \frac{e^2 \theta}{4 \pi^2} \int d t \,\, d^3 r \bE \cdot \bB. 
\eeq
The parameter $\theta$ is equal to $\pi$ in 3D TI, the only nonzero value, compatible with TR symmetry~\cite{Qi10}. 
The $\theta$-term in Eq.~\eqref{eq:31}, however, is a full derivative, and thus has no effect on Maxwell's equations in the bulk of the TI. 
Its only real effect is to generate the half-quantized AHE on the sample surfaces in the presence of surface magnetization (e.g. due to magnetic 
impurities). 

When TR is broken in the bulk, as it is in the Weyl semimetal state, a $\theta$-term with non-constant $\theta$ is now allowed by symmetry
\beq
\label{eq:32}
S_{\theta} = \frac{e^2}{32 \pi^2} \int d t \,\, d^3 r \theta(\br) \epsilon^{\mu \nu \alpha \beta} F_{\mu \nu} F_{\alpha \beta}.
\eeq
The only functional form, compatible with translational symmetry, is linear, i.e.
\beq
\label{eq:33}
\theta(\br) = 2 \bb \cdot \br, 
\eeq
where $\bb$ is a vector, which is odd under TR and even under I and coincides, as will be seen explicitly below, with 
the bulk spin splitting. 
If, in addition to TR, I symmetry is violated as well, the ``axion field" $\theta$ may also acquire a linear time dependence
\beq
\label{eq:34}
\theta(\br, t) = 2 \bb \cdot \br - 2 b_0 t, 
\eeq
where $b_0$ is the energy difference between the Weyl nodes, allowed if $I$ is violated. 
Unlike the $\theta$-term in 3D TI, Eq.~\eqref{eq:32} does modify Maxwell's equations in the bulk of the Weyl semimetal and thus 
has observable consequences, namely the already mentioned AHE and the Chiral Magnetic Effect (CME)~\cite{Kharzeev}, both of which will be discussed 
in detail below. 

The action of Eq.~\eqref{eq:32} may be described as being a consequence of chiral anomaly, an important concept in relativistic field theory~\cite{Adler69,Jackiw69,Nielsen83},
which has recently found its way into condensed matter physics and has been realized to play an important role in the theory of topologically-nontrivial 
states of matter~\cite{Ryu12,Nagaosa13,Wen13}.  
To gain a basic understanding of chiral anomaly, it is useful to go back to the generic low-energy model of Weyl semimetal, given by Eq.~\eqref{eq:3}. 
For simplicity, we will subsume the Fermi velocity $v_F$ in Eq.~\eqref{eq:3} into the definition of momentum and set the ``mass term" $\Delta = 0$. 
Then we obtain
\beq
\label{eq:35}
H = \tau^z \bsigma \cdot \bk + b_0  \tau^z + \bb \cdot \bsigma,  
\eeq
where we have included an energy difference between the Weyl nodes, described by the $b_0 \tau^z $ term, which is allowed when the I symmetry is violated, 
and oriented the TR-breaking vector $\bb$ along an arbitrary direction. 
It is convenient to represent the system in terms of an imaginary time action, including possible coupling of the electrons to an electromagnetic field
\beq
\label{eq:36}
S = \int d \tau d^3 r \,\,\psi^\dg \left[ \partial_{\tau} + i e A_0 + b_0 \tau^z + \tau^z \bsigma \cdot \left( -i \bnabla + e \bA + \bb \tau^z \right) \right] \psi^\pdg,
\eeq
where $A_{\mu} = (A_0, \bA)$ is the electromagnetic gauge potential and $\psi^\dg, \psi^\pdg$ are the 4-component spinor Grassman fields. We have suppressed all 
explicit spinor indices in the Grassmann fields for brevity. 
We now observe that in addition to the charge conservation symmetry, the imaginary time action Eq.~\eqref{eq:36} possesses an extra chiral symmetry
\beq 
\label{eq:37}
\psi \rightarrow e^{-i \tau^z \theta/2} \psi, 
\eeq 
which expresses an apparent separate conservation of the number of fermions of left and right chirality. This suggests that the terms $b_0 \tau^z$ and $\bb \cdot \bsigma$ in 
Eq.~\eqref{eq:36} could be eliminated by a gauge transformation
\beq 
\label{eq:38}
\psi \rightarrow e^{- i \tau^z \theta(\br, \tau)/2} \psi,\,\, \psi^\dg \rightarrow \psi^\dg e^{i \tau^z \theta(\br, \tau)/2},
\eeq
where $\theta(\br, \tau) = 2 \bb \cdot \br - 2 i b_0 \tau$ and one should keep in mind that $\psi$ and $\psi^\dg$ are not complex conjugates of each other, but are independent variables in the fermion path integral. 
The imaginary time action then becomes
\beq
\label{eq:39}
S = \int d \tau d \br \,\, \psi^\dg \left[ \partial_{\tau} + i e A_0 
 + \tau^z \bsigma \cdot \left( -i \bnabla + e \bA \right) \right] \psi^\pdg, 
\eeq
which describes two Weyl nodes of opposite chirality, existing at the same point in momentum space and in energy. 
This argument then leads one to the conclusion that the system of Weyl nodes, separated in energy and momentum, is equivalent 
to the system of two degenerate Weyl nodes and thus does not possess any special transport properties, which is incorrect. 
What is missing in the above argument is exactly the chiral anomaly: while the imaginary time action Eq.~\eqref{eq:36} does indeed 
possess the chiral symmetry, the gauge transformation of Eq.~\eqref{eq:38} changes not only the action itself, but also the measure of the path integral. 
This change of the path integral measure is what gives rise to the $\theta$-term Eq.~\eqref{eq:32}. 
This may be shown explicitly using the Fujikawa's method~\cite{Fujikawa,Zyuzin12}. 
We refer the reader to Ref.~\cite{Zyuzin12} for technical details. 

To see the physical consequences of the $\theta$-term, we integrate Eq.~\eqref{eq:32} by parts and eliminate a total derivative term. 
This leads to the following action, which resembles the Chern-Simons term in $2+1$-dimensions
\beq
\label{eq:40}
S_{\theta} = - \frac{e^2}{8 \pi^2} \int dt d^3 r \,\, \partial_{\mu} \theta \epsilon^{\mu \nu \alpha \beta} A_{\nu} \partial_{\alpha} A_{\beta}. 
\eeq
Varying Eq.~\eqref{eq:40} with respect to the vector potential, we obtain the following expression for the current
\beq
\label{eq:41}
j_{\nu} = \frac{e^2}{2 \pi^2} b_{\mu} \epsilon^{\mu \nu \alpha \beta} \partial_{\alpha} A_{\beta}, \,\,\, \mu = 1, 2, 3, 
\eeq
and 
\beq
\label{eq:42}
j_{\nu} = - \frac{e^2}{2 \pi^2} b_0 \epsilon^{0 \nu \alpha \beta} \partial_{\alpha} A_{\beta}. 
\eeq
Eq.~\eqref{eq:41} clearly represents the AHE,
while Eq.~\eqref{eq:42} represents another effect, related to chiral anomaly in Weyl semimetals, the Chiral Magnetic Effect~\cite{Kharzeev}, whose 
physical meaning is somewhat subtle and will be discussed in detail later. 

However, the picture of chiral anomaly and related transport phenomena, presented above, while simple and appealing, is not fully satisfactory, for the following 
reasons. 
Chiral anomaly is a sharply-defined concept in the context of relativistic field theory, where 
massless fermions in unbounded momentum space possess exact chiral symmetry Eq.~\eqref{eq:37}, which may be violated by the anomaly. 
In the condensed matter context, however, the situation is less clear. Even though chiral symmetry may be formally defined in a low-energy model of a Weyl semimetal 
Eq.~\eqref{eq:35}, in which the band 
dispersion is approximated as being exactly linear and unbounded, no real microscopic model of Weyl semimetal actually possesses such an exact symmetry. 
It may only appear as an approximate low-energy symmetry. 
Since the chiral symmetry is not present to begin with, it is then unclear to what extent is it meaningful to speak of its violation by chiral anomaly. 
The role of impurity scattering, present in any real condensed matter system, and important even conceptually in any discussion of transport phenomena; as well as the role of 
finite electron or hole density, present when the Fermi energy does not exactly coincide with the location of the Weyl nodes, 
are also completely unclear in the above field-theory discussion. 
In the remainder of this paper we will thus develop a fully microscopic theory of chiral anomaly and its manifestations, including the contribution of impurity scattering, and finite 
charge carrier density, based on the TI-NI heterostructure model, described in Section~\ref{sec:2}, which does not possess chiral symmetry, yet exhibits both the AHE and CME, described above. 

\subsection{Microscopic theory: Anomalous Hall Effect}
\label{sec:3.2}
We will now go back to our microscopic model of a Weyl semimetal, described in Section~\ref{sec:2}. 
We want to explicitly derive the topological term in the action of the electromagnetic field by integrating out 
electron variables, coupled to the field. 
We then start from the imaginary time action of electrons in the Weyl semimetal, coupled to electromagnetic field
\beq
\label{eq:43}
S = \int d \tau d^3 r \left\{ \Psi^\dg(\br, \tau) \left[\partial_{\tau} - \mu + i e A_0 (\br, \tau) + \hat H\right] \Psi^\pdg(\br, \tau)\right\}, 
\eeq
where $A_0(\br, \tau)$ is the scalar potential and 
\beq
\label{eq:44}
\hat H = v_F \tau^z (\hat z \times \bsigma) \cdot \left(- i \bnabla + e \bA \right) + \hat \Delta + b \sigma^z,
\eeq
is the Hamiltonian of noninteracting electrons in Weyl semimetal, minimally coupled to the vector potential $\bA$. 
We will ignore the $z$-component of the vector potential as it will not play any role in what follows. 

Using the results of Section~\ref{sec:2} for the eigenvalues and eigenvectors of the Weyl semimetal Hamiltonian in the absence of the electromagnetic 
field, we can now integrate out electron variables in Eq.~\eqref{eq:43} and obtain an effective action for the electromagnetic field, induced by coupling to the electrons. 
This action will contain two distinct kinds of contributions. The first kind will contain terms, proportional to $\bE^2$ and $\bB^2$, 
where $\bE$ and $\bB$ are the electric and magnetic fields. These terms describe the ordinary electric and magnetic polarizability of the material, and we will not discuss this 
part of the response here. 
The second kind contains the ``topological" contribution, which has 
the form of the ``3D Chern-Simons" term of Eq.~\eqref{eq:40}. 
Adopting the Coulomb gauge $\bnabla \cdot \bA = 0$, we obtain
\beq
\label{eq:45}
S = \sum_{\bq, i\Omega}\epsilon^{z 0 \alpha \beta} \Pi(\bq, i\Omega) A_0(-\bq, -i\Omega) \hat q_{\alpha} A_{\beta}(\bq, i \Omega),
\eeq 
where $\hat q_{\alpha} = q_{\alpha}/ q$ and summation over repeated indices is implicit. The $z$-direction in Eq.~\eqref{eq:45} is picked out by the 
magnetization $b$. 
The response function $\Pi(\bq, i\Omega)$ is given by
\beqa
\label{eq:46}
\Pi(\bq, i \Omega)&=&\frac{i e^2 v_F}{V}  \sum_{\bk} \frac{n_F[\xi_{s' t'}( \bk)] - n_F[\xi_{s t} (\bk + \bq)]}{i \Omega + \xi_{s' t'}(\bk) - \xi_{s t} (\bk + \bq)} \nonumber \\
&\times&\langle z^{s t}_{\bk + \bq} | z^{s' t'}_{\bk} \rangle \langle z^{s' t'}_{\bk}| \bsigma \cdot \hat q | z^{s t}_{\bk + \bq} \rangle,
\eeqa
where $\xi_{s t}(\bk) = \epsilon_{s t}(\bk) - \epsilon_F$ and summation over repeated band indices $s, t$ is implicit.  

To evaluate $\Pi(\bq, i\Omega)$ explicitly it is convenient to rotate coordinate axes so that $\bq = q \hat x$ and assume that $\epsilon_F > 0$. The $\epsilon_F < 0$ result is 
evaluated analogously. 
As seen from Eq.~\eqref{eq:46}, there exist two kinds of contributions to the response function $\Pi(\bq, i \Omega)$: {\em interband}, with $s \neq s'$, 
and {\em intraband}, with $s = s' = +$. 
Both in general contribute and need to be taken into account. 
Let us first evaluate the interband contributions. 
In this case we can set $i \Omega = 0$ in the denominator of Eq.~\eqref{eq:46} from the start.
Expanding to first order in $q$ and integrating over the transverse momentum components $k_{x,y}$, we obtain
\beqa
\label{eq:52}
&&\Pi^{inter}(\bq, i\Omega) = \frac{e^2 q}{8 \pi^2} \sum_t  \int_{-\pi/d}^{\pi/d} d k_z \,\, \textrm{sign} [m_t(k_z)] \nonumber \\
&\times& \left\{1 - \left[1 - \frac{|m_t(k_z)|}{\epsilon_F} \right] \Theta(\epsilon_F - |m_t(k_z)|)\right\}, 
\eeqa
where $\Theta(x)$ is the Heaviside step function. 
As clear from Eq.~\eqref{eq:52}, $\Pi^{inter}(\bq, i\Omega)$ consists of two distinct terms.  
The first term inside the curly brackets in Eq.~\eqref{eq:52} is the contribution of the completely filled $s = -$ bands. The second term is the contribution 
of incompletely filled $s = +$ bands. 

To understand the meaning of Eq.~\eqref{eq:52}, let us now evaluate the {\em intraband} contribution to $\Pi(\bq, i \Omega)$. 
In this case we have $s = s' = +$ in Eq.~\eqref{eq:46}. It is then clear that, unlike in the case of the interband contribution, evaluated 
above, the value of the intraband contribution depends on the order in which the limits of $\bq  \rightarrow 0$ and $\Omega \rightarrow 0$ 
are taken. If the limit is taken so that $\Omega/ v_F |\bq| \rightarrow \infty$, then one is evaluating the DC limit of a transport quantity, i.e the optical Hall 
conductivity. In this case, the intraband contribution vanishes identically. 
On the other hand, if the limit is taken so that $\Omega/ v_F |\bq| \rightarrow 0$, then one is evaluating an equilibrium thermodynamic property, whose 
physical meaning will become clear below. 
In this case, the intraband contribution is not zero and is given by 
\beqa
\label{eq:59}
\Pi^{intra}(\bq, i\Omega) = \frac{i e^2 v_F}{V}\sum_t  \sum_{\bk} \left.\frac{d n_F(x)}{d x} \right|_{x=\epsilon_t(\bk) - \epsilon_F} \langle z^{+ t}_{\bk + \bq} | z^{+ t}_{\bk} \rangle \langle z^{+t}_{\bk} |\bsigma \cdot \hat q| z^{+ t}_{\bk + \bq} \rangle. \nonumber \\
\eeqa
The derivative of the Fermi distribution function in Eq.~\eqref{eq:59} expresses the important fact that $\Pi^{intra}(\bq, i \Omega)$ is associated 
entirely with the Fermi surface, unlike $\Pi^{inter}(\bq, i\Omega)$, to which all filled states contribute, including states on the Fermi surface.  
Expanding to first order in $q$ as before and evaluating the integrals over $k_{x,y}$, we obtain
\beq
\label{eq:61}
\Pi^{intra}(\bq, i\Omega) = - \frac{e^2 q}{8 \pi^2} \sum_t \int_{-\pi/d}^{\pi/d} d k_z \, \, \frac{m_t(k_z)}{\epsilon_F} \Theta(\epsilon_F - |m_t(k_z)|).
\eeq
i.e. the intraband contribution to $\Pi(\bq, i\Omega)$ is equal to the second term in the square brackets in Eq.~\eqref{eq:52} in magnitude, 
but opposite in sign. 
Combining the inter- and intraband contributions to $\Pi(\bq, i\Omega)$ we thus obtain 
\beq
\label{eq:62}
\Pi(\bq, i\Omega) = \frac{e^2 q}{8 \pi^2} \sum_t \int_{-\pi/d}^{\pi/d} d k_z \,\, \textrm{sign} [m_t(k_z)]  \left[1 - \Theta(\epsilon_F - |m_t(k_z)|)\right],
\eeq
i.e. the last term in Eq.~\eqref{eq:52} cancels out when 
the low-frequency, long-wavelength limit is taken in such a way that $\Omega/v_F |\bq| \rightarrow 0$. 
On the other hand, when $\Omega/ v_F |\bq| \rightarrow \infty$, the intraband contribution vanishes and 
\beq
\label{eq:63}
\Pi(\bq, i\Omega) = \Pi^{inter}(\bq, i\Omega).
\eeq
This physical difference in the kind of response the system exhibits is the basis of Streda's separation of contributions to the Hall conductivity into $\sigma_{xy}^I$ and 
$\sigma_{xy}^{II}$~\cite{Streda}.
Our analysis makes it clear that this is the most physically-meaningful separation of contributions to the intrinsic anomalous Hall conductivity, since it corresponds 
to distinct and, at least in principle, separately measurable, contributions to the response function $\Pi(\bq, i\Omega)$. 

\begin{figure}[t]
\subfigure[]{
   \label{fig:2a}
  \includegraphics[width=7cm]{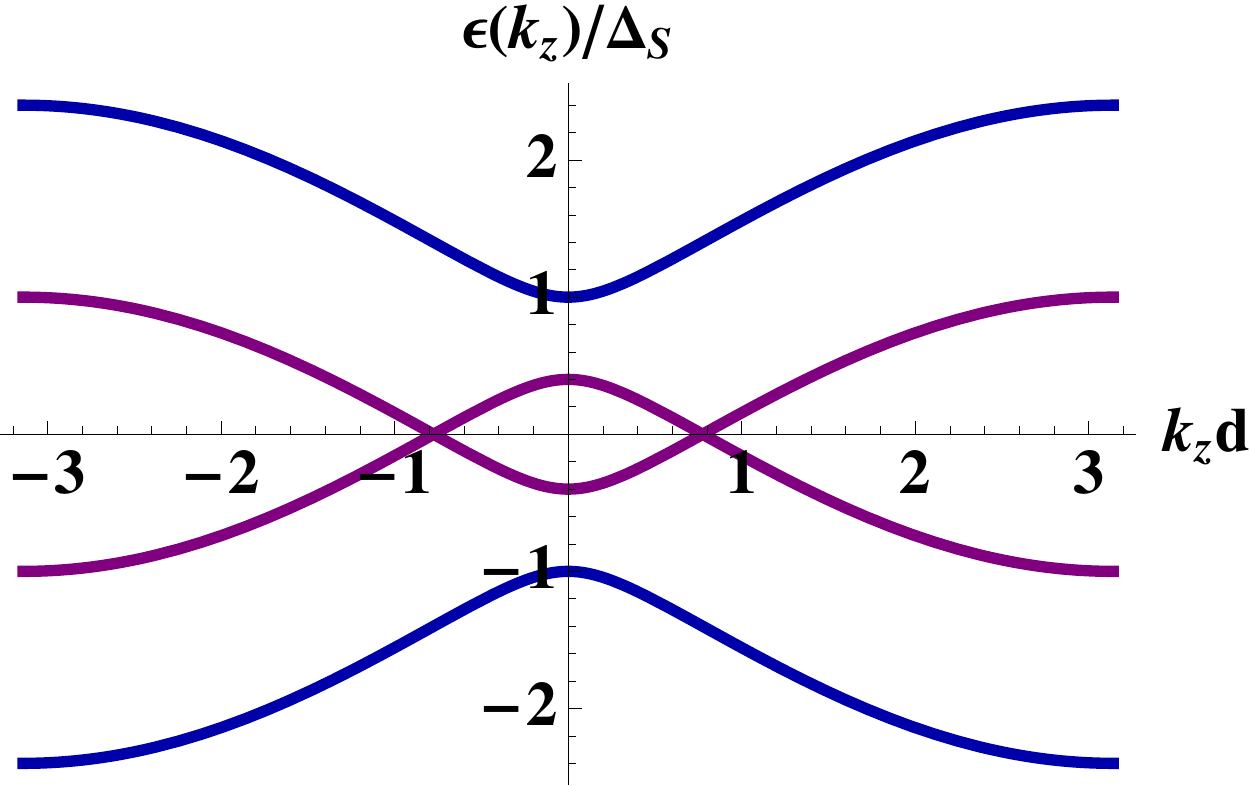}}
\subfigure[]{
  \label{fig:2b}
   \includegraphics[width=7cm]{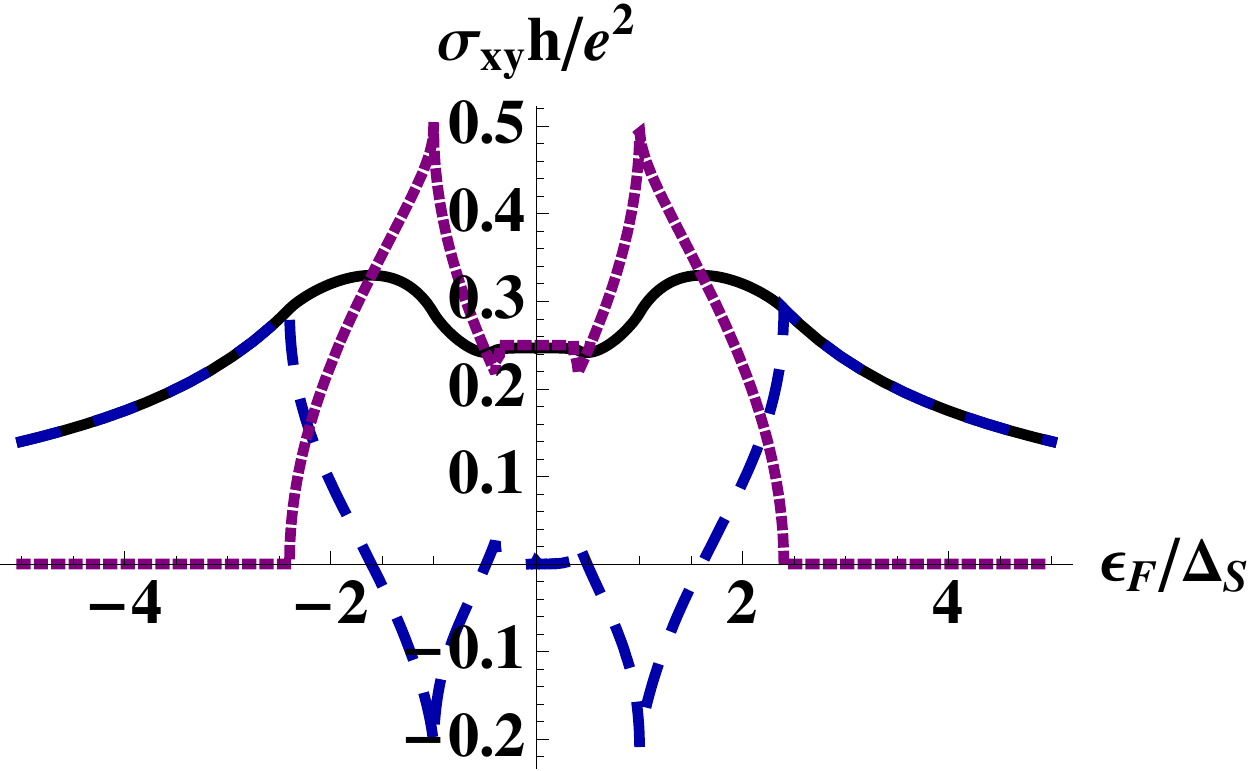}}
  \caption{(a) Plot of the band edges along the $z$-direction in momentum space.
  The parameters are such that two Weyl nodes are present.
  (b) Total intrinsic anomalous Hall conductivity (solid line), $\sigma_{xy}^I$ (dashed line), and $\sigma_{xy}^{II}$ (dotted line). Note that the van Hove-like singularities in 
  $\sigma^I_{xy}$ and $\sigma^{II}_{xy}$, associated with band edges, mutually cancel and the total Hall conductivity $\sigma_{xy}$ is a smooth function of the Fermi energy.}
    \label{fig:2}
\end{figure}
\begin{figure}[t]
\subfigure[]{
   \label{fig:3a}
  \includegraphics[width=7cm]{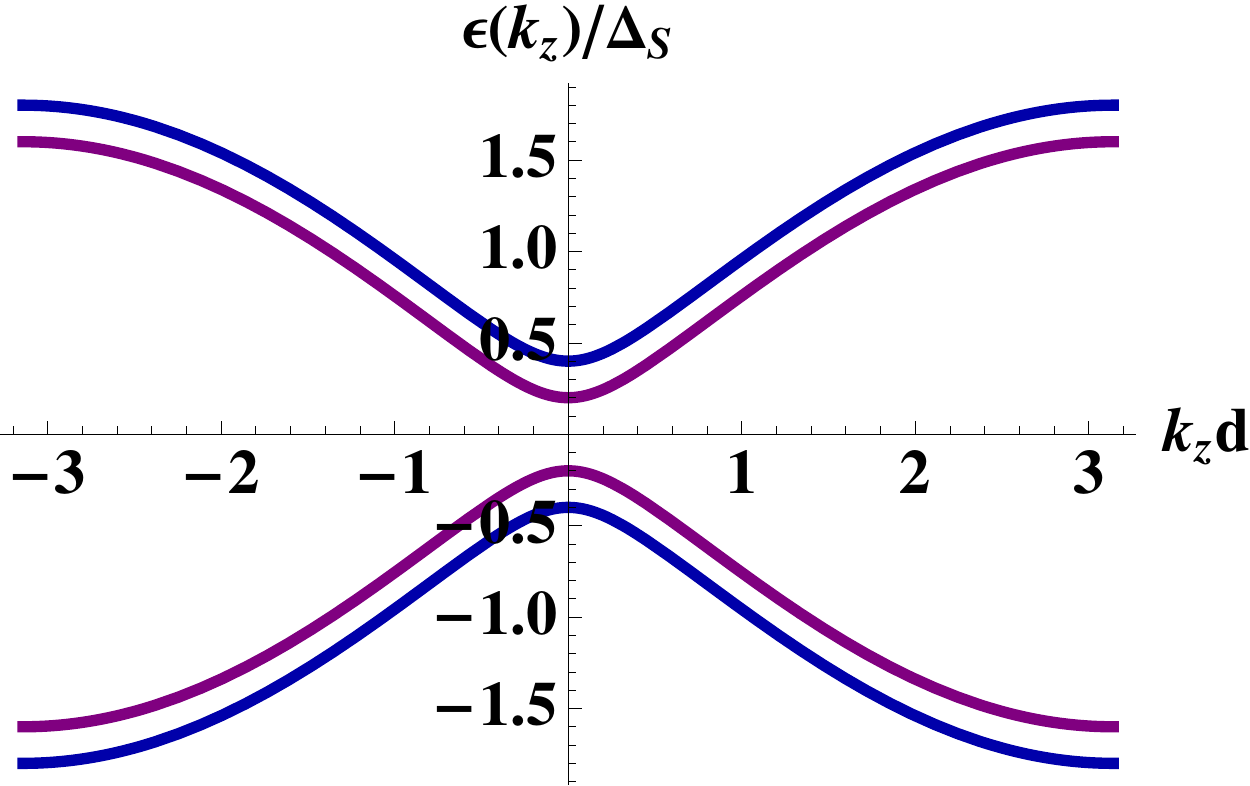}}
\subfigure[]{
  \label{fig:3b}
   \includegraphics[width=7cm]{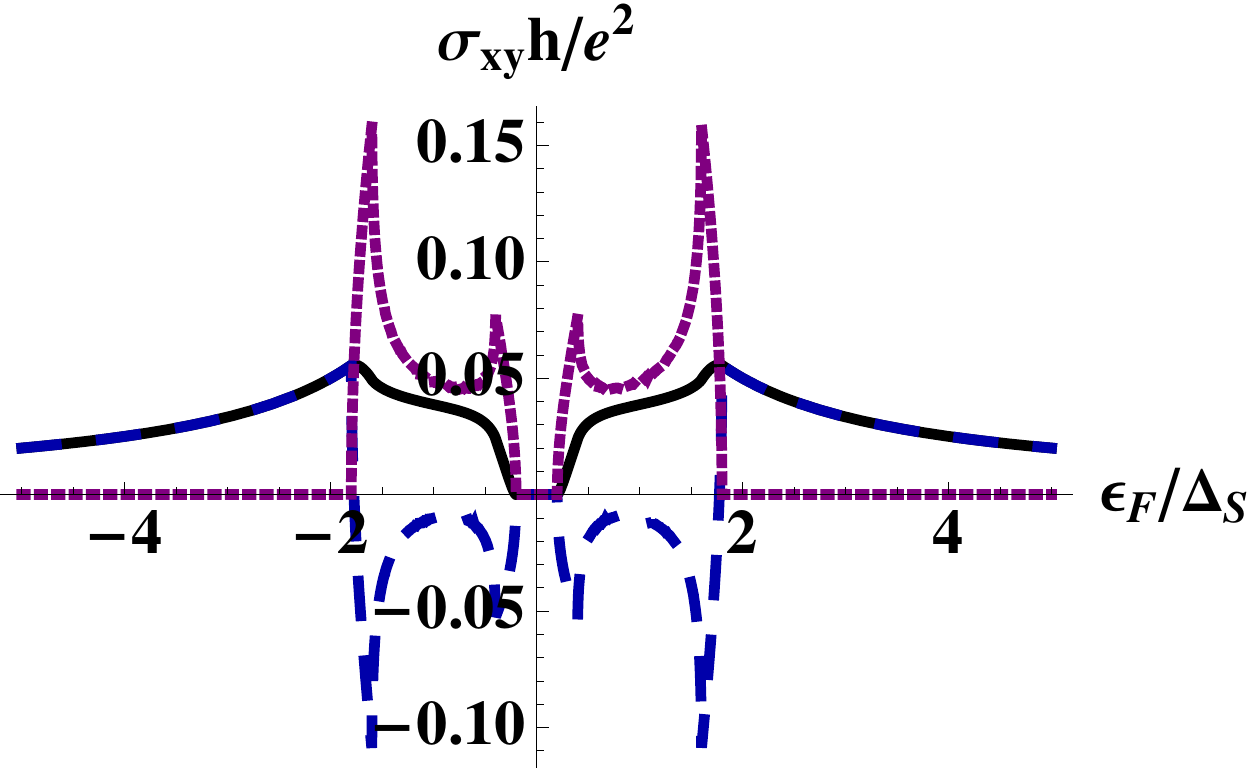}}
  \caption{(a) Plot of the band edges along the $z$-direction in momentum space. The spin splitting is not large enough for the 
  Weyl nodes to appear (i.e. $b < b_{c1}$) and the spectrum has a full gap. 
  (b) Total intrinsic anomalous Hall conductivity (solid line), $\sigma_{xy}^I$ (dashed line), and $\sigma_{xy}^{II}$ (dotted line).}
    \label{fig:3}
\end{figure}

Generalizing the above results to arbitrary sign of $\epsilon_F$ we finally obtain
\beq
\label{eq:64}
S = - i \epsilon^{z 0 \alpha \beta} \sigma_{xy} \int d^3 r d \tau A_0(\br, \tau) \partial_{\alpha} A_{\beta}(\br, \tau), 
\eeq
where, if the low-frequency long-wavelength limit is taken so that $\Omega/ v_F |\bq| \rightarrow \infty$:
\beqa
\label{eq:65}
\sigma_{xy}&=&\frac{e^2}{8 \pi^2} \sum_t \int_{-\pi/d}^{\pi/d} d k_z \,\, \left\{\textrm{sign} [m_t(k_z)] \left[\Theta(\epsilon_F + |m_t(k_z)|) - \Theta(\epsilon_F - |m_t(k_z)|)\right] \right.\nonumber \\
&+&\left. \frac{m_t(k_z)}{|\epsilon_F|} \Theta(|\epsilon_F| - |m_t(k_z)|) \right\}.
\eeqa
This expression corresponds to the full DC anomalous Hall conductivity. 
On the other hand, when the low-frequency, long-wavelength limit is taken so that $\Omega/ v_F |\bq| \rightarrow 0$, we obtain
\beqa
\label{eq:66}
\sigma_{xy}&=&\sigma_{xy}^{II}=\frac{e^2}{8 \pi^2} \sum_t \int_{-\pi/d}^{\pi/d} d k_z \,\, \textrm{sign} [m_t(k_z)] \nonumber \\
&\times&\left[\Theta(\epsilon_F + |m_t(k_z)|) - \Theta(\epsilon_F - |m_t(k_z)|)\right].
\eeqa
This is precisely the Streda's $\sigma^{II}_{xy}$ contribution to the Hall conductivity, which is a thermodynamic equilibrium quantity, 
equal to
\beq
\label{eq:67}
\sigma^{II}_{xy} = e \left(\frac{\partial N}{\partial B}\right)_{\mu}, 
\eeq
where $N$ is the total electron number. This relation follows immediately from Eq.~\eqref{eq:64} and the order of limits
$\Omega/ v_F |\bq| \rightarrow 0$, which corresponds to thermodynamic equilibrium.
Correspondingly, the $\sigma^{I}_{xy}$ contribution is given by
\beq
\label{eq:68}
\sigma^I_{xy} = \sigma_{xy} - \sigma^{II}_{xy} =  \frac{e^2}{8 \pi^2} \sum_t \int_{-\pi/d}^{\pi/d} d k_z \frac{m_t(k_z)}{|\epsilon_F|} \Theta(|\epsilon_F| - |m_t(k_z)|).
\eeq
As clear from the above analysis, $\sigma^I_{xy}$ is the contribution to $\sigma_{xy}$ that can be associated with 
states on the Fermi surface. This contribution is nonuniversal, i.e. it depends on details of the electronic structure and 
is a continuous function of the Fermi energy. 
$\sigma^{II}_{xy}$, on the other hand, is the contribution of all states below the Fermi energy and is a thermodynamic 
equilibrium property of the ferromagnet. It attains a universal value,  which depends only on the distance between the 
Weyl nodes, when the Fermi energy coincides with the nodes, i.e. when $\epsilon_F = 0$
\beq
\label{eq:69}
\sigma_{xy}^{II} = \frac{e^2 {\cal K}}{4 \pi^2},
\eeq
where 
\beq
\label{eq:70}
{\cal K} = \frac{2}{d} \arccos\left(\frac{\Delta_S^2 + \Delta_D^2 - b^2}{2 \Delta_S  \Delta_D} \right), 
\eeq
is the distance between the Weyl nodes. 
When $b > b_{c2}$, the Weyl nodes annihilate at the edges of the Brillouin zone and a gap opens up. 
In this case ${\cal K} = 2 \pi/d$, i.e. a reciprocal lattice vector and $\sigma^{II}_{xy}$ is quantized as long as the Fermi level is in the gap~\cite{Halperin92}.
Both contributions, along with the total anomalous Hall conductivity $\sigma_{xy}$ are plotted as a function of the Fermi energy in Figs.~\ref{fig:2},~\ref{fig:3}
in two different cases: when Weyl nodes are present and when they are not. 
The former occurs when $b_{c1} < b < b_{c2}$. 
As can be seen from Fig.~\ref{fig:2}, Weyl nodes provide the dominant contribution to $\sigma^{II}_{xy}$ and to the total Hall conductivity $\sigma_{xy}$, if the 
Fermi level is not too far from the nodes. 

It is interesting to note the following property, which is evident from Fig.~\ref{fig:2}. 
Both the total anomalous Hall conductivity $\sigma_{xy}$ and the two distinct contributions to it, $\sigma_{xy}^{I,II}$ appear to exhibit a 
quasi-plateau behavior when $\epsilon_F$ is not too far from the Weyl nodes. To understand the origin of this behavior, consider the derivative 
of $\sigma^{II}_{xy}$ with respect to the Fermi energy
\beq
\label{eq:71}
\frac{\partial \sigma^{II}_{xy}}{\partial \epsilon_F} = - \frac{e^2}{8 \pi^2} \sum_t \int_{-\pi/d}^{\pi/d} d k_z \textrm{sign}[m_t(k_z)] \delta(\epsilon_F - |m_t(k_z)). 
\eeq
This is straightforward to evaluate analytically and  we obtain
\beqa
\label{eq:72}
\frac{\partial \sigma_{xy}^{II}}{\partial \epsilon_F}&=&- \frac{e^2}{8 \pi^2} \int_{-\pi/d}^{\pi/d} d k_z \left[\delta(\Delta(k_z) - b + \epsilon_F) - \delta(\Delta(k_z) - b -\epsilon_F) \right] \nonumber \\ &=&\frac{e^2}{4 \pi^2}\left(1/\tilde v_{F+} - 1/\tilde v_{F-} \right), 
\eeqa
where we have assumed that $\epsilon_F$ is sufficiently close to zero, so that only the $t = -$ bands contribute to the integral, and 
\beq
\label{eq:73}
\tilde v_{F\pm} = \frac{d}{2 (b \pm \epsilon_F)} \sqrt{[(b \pm \epsilon_F)^2 - b_{c1}^2] [b_{c2}^2 - (b \pm \epsilon_F)^2]}, 
\eeq
are the two Fermi velocities, corresponding to two pairs of solutions of the equation $|b - \Delta(k_z)| = \epsilon_F$, which 
arise from the Fermi level crossing the $s=+, t=-$ band on the two sides of each Weyl node along the $z$-axis in momentum space. 
The two Fermi velocities are nearly equal when the band dispersion near the nodes is almost perfectly linear, but start to differ significantly 
when deviations from linearity become noticeable. 
Explicitly, as long as $b_{c1} \ll b \pm \epsilon_F \ll b_{c2}$, both Fermi velocities are independent 
of the Fermi energy and thus $\partial \sigma_{xy}^{II}/\partial \epsilon_F$ vanishes. 
A useful way to think about this is in terms of an approximate {\em chiral symmetry}, which emerges  in this system in the limit of small $\epsilon_F$, and which manifests 
in an almost perfectly linear band dispersion near the Weyl nodes. 
 
By a nearly identical calculation it is easy to show that $\partial \sigma^{I}_{xy} / \partial \epsilon_F$ also vanishes when $\epsilon_F$ is sufficiently close to zero. 
This is the origin of the quasi-plateau behavior in Fig.~\ref{fig:2}.
This result implies that the intrinsic anomalous Hall conductivity is equal to its thermodynamic equilibrium part, $\sigma^{II}_{xy}$, not just when the Fermi 
energy coincides with the Weyl nodes, but even away from them as long as the band dispersion may be assumed to be linear~\cite{Burkov14}.
This property will be discussed in more detail in Section~\ref{sec:4}. 

When attempting to understand the physical meaning of the two contributions to the anomalous Hall conductivity, $\sigma^I_{xy}$ and $\sigma^{II}_{xy}$, 
it may be tempting to say that, while $\sigma^{I}_{xy}$ is clearly associated with states on the Fermi surface, $\sigma^{II}_{xy}$ might perhaps be associated 
with the chiral Fermi arc edge states. Unfortunately, such an interpretation is clearly incorrect, at least in the context of the present model, since, as can be 
seen in Fig.~\ref{fig:3}, $\sigma^{II}_{xy}$ may be nonzero even when Weyl nodes and thus the Fermi arc edge states are absent. More work is thus needed 
to fully understand the relation between $\sigma^{II}_{xy}$ and the edge states.

\subsection{Microscopic theory: Chiral Magnetic Effect}
\label{sec:3.3}
In this subsection we consider the second part of the topological response in Weyl semimetals, namely the CME. 
This arises due to the presence of an energy difference between the Weyl nodes, as in Eq.~\eqref{eq:34}. 
To induce such an energy difference in our microscopic model, we need to find an operator that acts as an ``axial chemical potential" term, 
shifting the nodes with different chirality in opposite directions in energy. 
The expression for this operator may be found based on symmetry considerations. 
We define an ``axial charge density" operator, $\hat n_a$, as a local operator, which is odd under inversion and $z \rightarrow -z$ reflections, even under 
time reversal, but odd under time reversal combined with rotation of the spin quantization axis by $\pi$ around either $x$ or $y$ axis. 
This leads to the following expression for the axial charge density operator
\beq
\label{eq:74}
\hat n_a = \tau^y \sigma^z. 
\eeq
Adding a term $- \mu_a \hat n_a$, where $\mu_a$ is the ``axial chemical potential", to the multilayer Hamiltonian, introduces 
an energy difference between the Weyl nodes of magnitude
\beq
\label{eq:75}
\Delta \epsilon = \frac{2 \mu_a \tilde v_F}{\Delta_S d}, 
\eeq
where 
\beq
\label{eq:76}
\tilde v_F = \frac{d}{2 b} \sqrt{(b^2 - b_{c1}^2)(b_{c2}^2 - b^2)}, 
\eeq
is the $z$-component of the Fermi velocity at the location of the Weyl nodes. 

Similarly to the previous subsection, we now couple the electrons to electromagnetic 
field and the axial chemical potential and integrate out the electron variables to obtain 
an induced action for the electromagnetic field. 
We will assume, without loss of generality, that the electromagnetic field consists of a magnetic field in the $z$-direction, 
and a vector potential $A_z$, whose time derivative gives the $z$-component of the electric field $E_z = - \partial_t A_z$. 
We will allow for a time and $z$-coordinate dependence of the vector potential $A_z$ and of the axial chemical potential $\mu_a$, but 
assume that the magnetic field is time-independent and uniform. 

For this calculation it is convenient to use the Landau level (LL) basis of the multilayer placed in a uniform external magnetic field along the growth 
direction, i.e. the basis of the eigenstates of the following Hamiltonian
\beq
\label{eq:77}
{\cal H}(k_z) = v_F \tau^z (\hat z \times \bsigma) \cdot \left(- i \bnabla + e \bA \right) + \hat \Delta(k_z) + b \sigma^z, 
\eeq
Adopting Landau gauge for the vector potential $\bA = x B \hat y$, we obtain the following expressions LL eigenstates, which have the form, typical for LLs in Dirac systems
\beq
\label{eq:78}
| n, a, k_y, k_z \rangle = \sum_{\tau} \left[z^a_{n \upa \tau}(k_z) | n -1, k_y, k_z, \upa, \tau \rangle + z^a_{n \da \tau}(k_z) | n, k_y, k_z, \da, \tau \rangle \right]. 
\eeq 
Here 
\beq
\label{eq:79}
\langle \br | n, k_y, k_z, \sigma, \tau \rangle = \frac{1}{\sqrt{L_z}} e^{i k_z z} \phi_{n k_y}(\br) | \sigma, \tau \rangle, 
\eeq
$\phi_{n k_y}(\br)$ are the Landau-gauge orbital wavefunctions, and $\sigma, \tau$ are the spin and pseudospin indices 
respectively. Finally, the four-component eigenvector $| z^a_{n}(k_z) \rangle$ may be written as a tensor product of the two-component spin and pseudospin eigenvectors, 
i.e. $| z^a_{n}(k_z) \rangle = | v^a_{n}(k_z) \rangle \otimes | u^a(k_z) \rangle$, where 
\beqa
\label{eq:80}
&&|v^{s t}_{n}(k_z) \rangle = \frac{1}{\sqrt{2}} \left(\sqrt{1 + s \frac{m_t(k_z)}{\epsilon_{n t}(k_z)}}, - i s \sqrt{1 - s \frac{m_t(k_z)}{\epsilon_{n t}(k_z)}} \right), \nonumber \\
&&|u^t(k_z) \rangle = \frac{1}{\sqrt{2}} \left(1, t \frac{\Delta_S + \Delta_D e^{- i k_z d}}{\Delta(k_z)} \right),
\eeqa
and the eigenstate energies are given by
\beq
\label{eq:80.1}
\epsilon_{n s t}(k_z) = s \sqrt{2 \omega_B^2 n + m_t^2(k_z)} \equiv s \epsilon_{n t}(k_z), 
\eeq
where $\omega_B = v_F/ \ell_B$ is the Dirac cyclotron frequency and $\ell_B = 1/\sqrt{e B}$ is the magnetic length. 

As in all Dirac systems, the lowest $n = 0$ LL is special and needs to be considered separately. The $s$ quantum number is absent in this case and  
taking $B > 0$ for concreteness, we have $\epsilon_{n t}(k_z) = - m_t(k_z)$, and $|v^t_{0}(k_z) \rangle = (0,1)$. 

The topological term, of interest to us, is proportional to the product of $\mu_a$ and $A_z$. 
Integrating out the electron variables and leaving only this term in the imaginary time action, we obtain
\beq
\label{eq:85}
S = B \sum_{q, i \Omega} \Pi(q, i\Omega) A_z(q, i\Omega) \mu_a(-q, -i \Omega)
\eeq
where the response function $\Pi(q, i\Omega)$ is given by:
\beqa
\label{eq:86}
\Pi(q, i\Omega)&=&\frac{e}{2 \pi L_z} \sum_{n, k_z} \frac{n_F[\xi_{n a'}(k_z)] - n_F[\xi_{n a}(k_z + q)]}{i \Omega + \xi_{n a'}(k_z) - \xi_{n a}(k_z + q)} \nonumber \\
&\times& \langle z^a_{n k_z}| \hat j_z(k_z) | z^{a'}_{n k_z} \rangle \langle z^{a'}_{n k_z} | \tau^y |z^a_{n k_z} \rangle. 
\eeqa
Here $n_F$ is the Fermi-Dirac distribution function, $\xi_{n a}(k_z) = \epsilon_{n a}(k_z) - \epsilon_F$, and the magnetic field $B$ in Eq.~\eqref{eq:85} arises 
from the Landau level orbital degeneracy. 
We have also ignored the $q$-dependence of the matrix elements in Eq.~\eqref{eq:86}, which is not important for small $q$. 

At this point we will specialize to the case of an undoped Weyl semimetal, i.e. set $\epsilon_F = 0$. 
Then it is clear from Eq.~\eqref{eq:86} that for Landau levels with $n \geq 1$, only terms with $s \neq s'$ contribute due to the difference of Fermi factors 
in the numerator. 
We are interested ultimately in the zero frequency and zero wavevector limit of the response function $\Pi(q, i\Omega)$. 
As already seen in the previous section in the context of AHE, the value of $\Pi(0,0)$ depends on the order in which the zero frequency and zero wavevector 
limits are taken.
Below we will consider both possibilities separately and discuss their physical meaning. 

Before we proceed with an explicit evaluation of the $q \rightarrow 0$ and $\Omega \rightarrow 0$ limit, let us note 
an important property of the response function $\Pi(q, \Omega)$.  
If we take into account the following symmetry properties of the matrix elements in Eq.~\eqref{eq:86}
\beqa
\label{eq:88}
&&\langle v^{+ t}_{n k_z} | v^{- t'}_{n k_z} \rangle  = - \langle v^{+ t'}_{n k_z} | v^{- t}_{n k_z} \rangle, \nonumber \\
&&\langle v^{+ t}_{n k_z} | \sigma^z| v^{- t'}_{n k_z} \rangle  =  \langle v^{+ t'}_{n k_z} | \sigma^z| v^{- t}_{n k_z} \rangle, 
\eeqa
it is easy to see that when the limit $\Omega \rightarrow 0$ is taken, independently of the value of $q$, the $n \geq 1$ Landau levels in fact 
do not contribute at all, mutually cancelling due to Eq.~\eqref{eq:88}. It follows that $\Pi(q, i\Omega)$ at small $\Omega$ is
determined completely by the contribution of the two $n = 0$ Landau levels, whose energy eigenvalues and the corresponding 
eigenvectors are independent of the magnetic field. 
This means that in the small $\Omega$ limit $\Pi(q, i \Omega)$ becomes independent of the magnetic field and the effective action in Eq.~\eqref{eq:85} 
then depends linearly on both $\mu_a$ and $B$, independently of the magnitude of $B$. 

Let us now proceed to explicitly evaluate $\Pi(0,0)$, which determines the low-frequency and long-wavelength 
response of our system. 
Let us first look at the situation when we send $q$ to zero before sending $\Omega$ to zero. 
Explicitly evaluating the matrix elements in Eq.~\eqref{eq:86} and the momentum integrals we obtain the following simple expression for $\Pi(0,0)$
\beq
\label{eq:93} 
\Pi(0,0) = - \frac{e^2}{4 \pi^2} \frac{2 \tilde v_F}{\Delta_S d}.
\eeq
Thus, after Wick's rotation $\tau \rightarrow i t$, $\Delta \epsilon \rightarrow  - i \Delta \epsilon$, we finally obtain the following result for the electromagnetic field action
\beq
\label{eq:94}
S =  - \frac{e^2 \Delta \epsilon}{4 \pi^2} \frac{2 \tilde v_F}{\Delta_S d} B \int d^3 r d t \,\, A_z(\br, t) \mu_a(\br, t),
\eeq
which, taking into account Eq.~\eqref{eq:75}, has precisely the form of Eq.~\eqref{eq:40}. 
Functional derivative of Eq.~\eqref{eq:94} with respect to $A_z$ gives the current that flows in response to magnetic 
field and axial chemical potential, i.e. the CME
\beq
\label{eq:95}
j_z  = - \frac{e^2 \Delta \epsilon}{4 \pi^2} B. 
\eeq

The physical interpretation of Eq.~\eqref{eq:95} requires some care. 
The issue is again the order of limits when calculating $\Pi(0,0)$, which always arises when calculating 
response functions in gapless systems.
Eqs.~\eqref{eq:94},\eqref{eq:95} was obtained by sending $q \rightarrow 0$ before $\Omega \rightarrow 0$. 
Let us now see what happens when $\Omega$ is sent to zero before taking the limit $q \rightarrow 0$. 
In this case, in addition to the contribution to $\Pi(0,0)$, given by Eq.~\eqref{eq:93}, which arises due to transitions 
between the $t = +$ and $t = -$ lowest ($n = 0$) Landau levels, there is an extra contribution due to the intra-Landau-level 
processes within the $t = -$ Landau level, which crosses the Fermi energy at the location of the Weyl nodes. 
This extra contribution is given by
\beq
\label{eq:96}
\tilde \Pi(0,0) = \frac{ e}{2 \pi L_z} \sum_{k_z} \left. \frac{d n_F(\epsilon)}{d \epsilon} \right|_{\epsilon = - m_-(k_z)} \langle z^-_{0 k_z} | \hat j_z(k_z) | z^-_{0 k_z} 
\rangle \langle z^-_{0 k_z} | \tau^y | z^-_{0 k_z} \rangle, 
\eeq
and is easily shown to be equal to Eq.~\eqref{eq:93} in magnitude, but opposite in sign, which means that in this case $\Pi(0,0)$ vanishes. 
Thus, the final result for  $\Pi(0,0)$ depends on the order in which the $q \rightarrow 0$ and $\Omega \rightarrow 0$ limits 
are taken
\beqa
\label{eq:97}
&&\lim_{\Omega \rightarrow 0} \lim_{q \rightarrow 0} \Pi(q, \Omega) = - \frac{e^2}{4 \pi^2} \frac{2 \tilde v_F}{\Delta_S d}, \nonumber \\
&&\lim_{q \rightarrow 0} \lim_{\Omega \rightarrow 0} \Pi(q, \Omega) = 0. 
\eeqa
What is the physical meaning of these two distinct orders of limits in calculating $\Pi(0,0)$? 
This is again identical to the AHE case, discussed in the previous section. 
When $q$ is taken to zero first, one is calculating the low-frequency limit of response to 
a time-dependent external field, in our case magnetic field along the $z$-direction. 
This response is finite and represents CME, described by Eq.~\eqref{eq:95}. 
If one takes $\Omega$ to zero first, however, one is calculating a thermodynamic property, in our 
case change of the ground state energy of the system in the presence of an additional static vector potential  
in the $z$-direction
\beq
\label{eq:98}
j_z = \frac{1}{V} \frac{\partial E(A_z)}{\partial A_z} = 0. 
\eeq
This could be nonzero in, for example, a current-carrying state of a superconductor, which 
possesses phase rigidity, but vanishes identically in our case~\cite{Franz13,Burkov13}. 

\section{Diffusive transport in Weyl metals}
\label{sec:4}
In the previous section we have focused on the basic properties of electromagnetic response 
of clean Weyl semimetals and metals (doped Weyl semimetals), neglecting the effect of impurity scattering. 
It is, however, always present, and while probably not of significant importance in an undoped Weyl semimetal
with density of states at the Fermi energy vanishing as $\sim \epsilon_F^2$, it may be expected to play an important role 
in a Weyl metal with $\epsilon_F$ significantly different from zero.  
If nothing else, impurity scattering is crucial for establishing a steady state under applied electric field, and thus must be 
included in any serious discussion of low-frequency transport phenomena. 
In this section we will thus generalize the theory of electromagnetic response, presented in Section~\ref{sec:3}, to the case of diffusive transport in dirty Weyl metals.
As will be demonstrated below, this generalization leads to two important new results. 

First, we will demonstrate that the anomalous Hall conductivity of a Weyl metal is purely intrinsic and universal, 
lacking both the so-called extrinsic contribution due to the impurity scattering and also the part of the intrinsic 
electronic-structure contribution, coming from incompletely filled bands. Only the purely intrinsic semi-quantized contribution, 
$\sigma_{xy}^{II}$, arising from completely filled bands and proportional to the separation between the Weyl nodes, survives 
in the Weyl metal, as long as the Fermi energy is close enough to the Weyl nodes, such that the band dispersion at the Fermi energy may be 
assumed to be linear. 

Second, we show that generalization of the chiral magnetic effect to the case of diffusive transport leads to a novel weak-field magnetoresistance 
effect, negative and quadratic in the magnetic field. We argue that this effect may be regarded as a universal smoking-gun transport characteristic 
of all Weyl metals. 

\subsection{Anomalous Hall effect in dirty Weyl metals}
\label{sec:4.1}
We start again from the heterostructure model of Weyl semimetal described in detail in the previous sections. 
We would like to evaluate the anomalous Hall conductivity of this model ferromagnetic Weyl metal, in the presence of impurity potential $V(\br) = V_0 \sum_a \delta(\br - \br_a)$, which we will assume for simplicity to be gaussian, with only second order correlators present: $\langle V(\br) V(\br ') \rangle = \gamma^2 \delta(\br - \br ')$, where $\gamma^2 = n_i V_0^2$ and $n_i$ is the impurity density.  
The higher-order correlators, which are known to be important for AHE in principle, as they give rise to the so-called skew-scattering contribution to the Hall conductivity, 
do not in fact affect our results in a significant way. 
We will also assume that the impurity potential is diagonal in both the pseudospin $\btau$ and the $t = \pm$ index, which labels the eigenstates of the
$\hat \Delta(k_z)$ operator Eq.~\eqref{eq:19}.  Again, this assumption 
is used only for computational simplicity and does not affect the essence of our results. 
To find the anomalous Hall conductivity, we will use the method of Section~\ref{sec:3.2}, which we find to be the most convenient one for our 
purposes, as it allows to most clearly separate physically distinct contributions to the AHE. 
Namely, we imagine coupling electromagnetic field to the electrons and integrating the electron degrees of freedom out to obtain an 
effective action for the electromagnetic field, which, at quadratic order, describes the linear response of the system. 
The part of this action we are interested in has the appearance of a Chern-Simons term, which, adopting the Coulomb gauge for the electromagnetic vector potential $\bnabla \cdot \bA = 0$, is given by
\beq
\label{eq:99}
S = \sum_{\bq, i\Omega}\epsilon^{z 0 \alpha \beta} \Pi(\bq, i\Omega) A_0(-\bq, -i\Omega) \hat q_{\alpha} A_{\beta}(\bq, i \Omega),
\eeq 
where $\hat q_{\alpha} = q_{\alpha}/ q$ and summation over repeated indices is implicit. The $z$-direction in Eq.~\eqref{eq:99} is again picked out by the 
magnetization $b$. As we will be interested specifically in the zero frequency and zero wavevector limits of the response function $\Pi(\bq, i \Omega)$, 
we will assume henceforth that $\bq = q \hat x$, which does not lead to any loss of generality due to full rotational symmetry in the $xy$-plane. 
The anomalous Hall conductivity is given by the zero frequency and zero wave vector limit of the response function $\Pi(\bq, i \Omega)$ as
\beq
\label{eq:100}
\sigma_{xy} = \lim_{i \Omega \rightarrow 0} \lim_{q \rightarrow 0} \frac{1}{q} \Pi(\bq, i \Omega).
\eeq
A significant advantage of Eq.~\eqref{eq:100}, compared to the more standard Kubo formula for the anomalous Hall conductivity, which relates 
it to the current-current correlation function, is that Eq.~\eqref{eq:100} ties the Hall conductivity to the response of a conserved quantity, 
i.e. the particle density. This means, in particular, that the response function $\Pi(\bq, i \Omega)$ must satisfy exact Ward identities, which follow from 
charge conservation, providing a useful correctness check on the results. 

The impurity average of the response function $\Pi(\bq, i\Omega)$ may be evaluated by the standard methods of diagrammatic perturbation theory. 
Due to our assumption that the impurity potential is diagonal in the band index $t$, we can do this calculation separately for each pair of bands, labeled 
by $t$, and then simply sum the individual contributions. We will thus omit the $t$ index in what follows, until we come to the final results. 
The retarded impurity averaged one-particle Green's functions have the following general form
\beq
\label{eq:101}
G^R_{\sigma_1 \sigma_2}(\bk, \epsilon) = \frac{z^s_{\bk \sigma_1} \bar z^s_{\bk \sigma_2}}{\epsilon - \xi^s_\bk + i /2 \tau_s}. 
\eeq
Here $s = \pm$, as before, labels the positive and negative energy pairs of bands (the sum over $s$ is made implicit above)
\beq
\label{eq:101.1}
\xi^s_\bk = s \epsilon_\bk - \epsilon_F = s \sqrt{v_F^2(k_x^2 + k_y^2) + m^2(k_z)} - \epsilon_F, 
\eeq
are the band energies, counted from the Fermi energy $\epsilon_F$, and 
\beq
\label{eq:101.2}
| z^s_\bk \rangle = \frac{1}{\sqrt{2}} \left(\sqrt{1 + s \frac{m(k_z)}{\epsilon_\bk}}, - i s e^{i \varphi} \sqrt{1 - s \frac{m(k_z)}{\epsilon_\bk}}\right), 
\eeq
is the corresponding eigenvector.
In what follows we will assume, for concreteness, that $\epsilon_F > 0$, i.e. the Weyl metal is electron-doped. 
The impurity scattering rates $1/\tau_{\pm}$ are given, in the Born approximation, by 
\beq
\label{eq:102}
\frac{1}{\tau_s(k_z)} = \frac{1}{\tau}  \left[1 +  s \frac{m(k_z) \langle m \rangle}{\epsilon_F^2} \right],
\eeq
where $1/\tau = \pi \gamma^2 g(\epsilon_F)$ and 
\beq
\label{eq:102.1}
g(\epsilon_F) = \int \frac{d^3 k}{(2 \pi)^3} \delta(\epsilon_\bk - \epsilon_F) = \frac{\epsilon_F}{4 \pi^2 v_F^2} \int_{-\pi/d}^{\pi/d} d k_z \Theta(\epsilon_F - |m(k_z)|), 
\eeq
is the density of states at Fermi energy. We have also defined the average of $m(k_z)$ over the Fermi surface as
\beq
\label{eq:103}
\langle m \rangle = \frac{1}{g(\epsilon_F)} \int \frac{d^3 k}{(2 \pi)^3} m(k_z) \delta(\epsilon_\bk - \epsilon_F). 
\eeq
\begin{figure}[t]
\begin{center}
  \includegraphics[width=12cm]{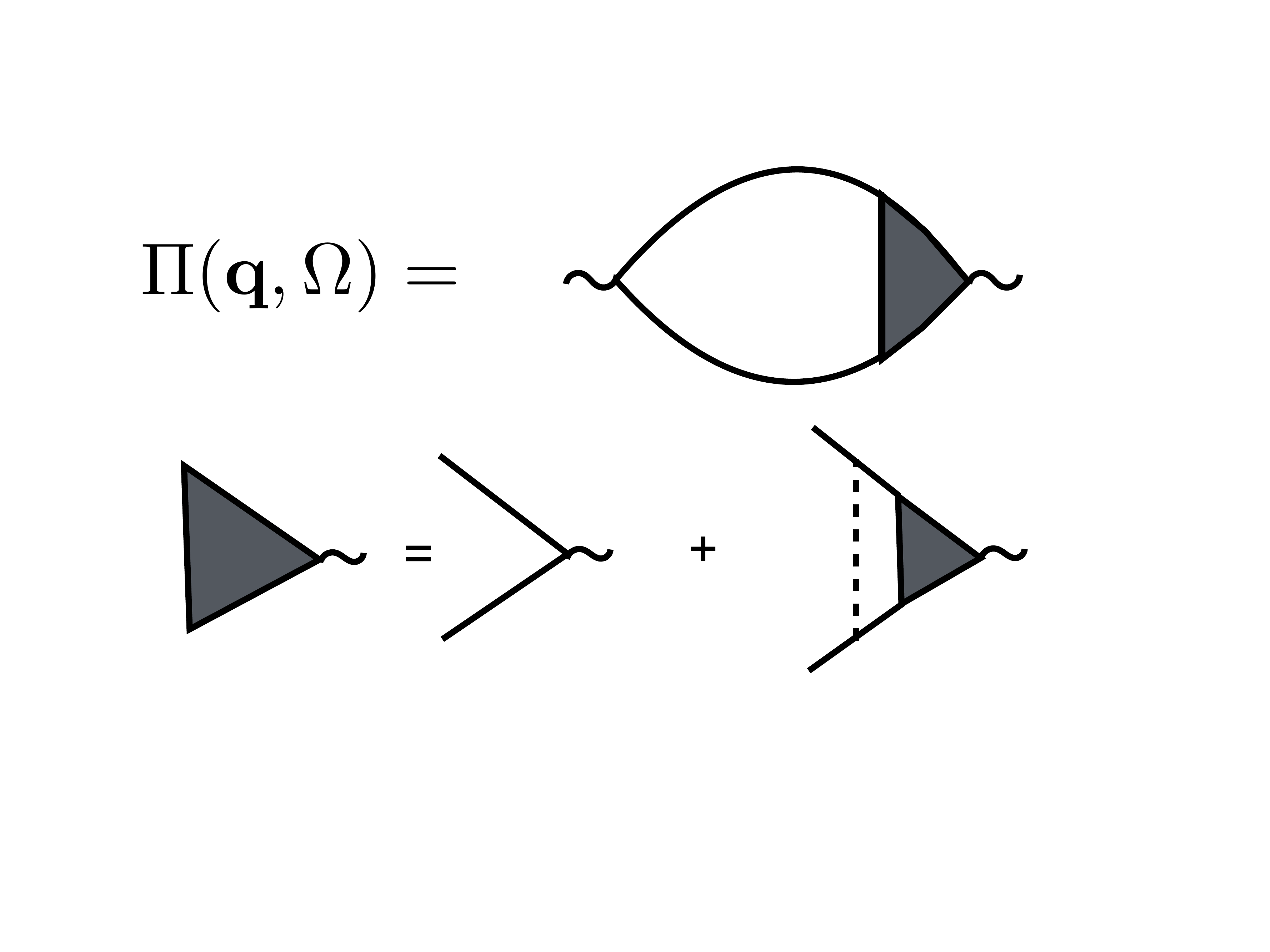}
\end{center}
\vspace{-2cm}
  \caption{Graphical representation of the impurity-averaged response function $\Pi(\bq, \Omega)$.}  
  \label{fig:3.5}
\end{figure} 
The impurity averaged response function, analytically continued to real frequency as $\Pi(\bq, \Omega) = \Pi(\bq, i \Omega \rightarrow \Omega + i \eta)$,
is given, in the self-consistent non-crossing approximation, by the sum of ladder diagrams, as illustrated in Fig.~\ref{fig:3.5}
\beq
\label{eq:104}
\Pi(\bq, \Omega) = \Pi^I(\bq, \Omega) + \Pi^{II}(\bq, \Omega), 
\eeq
where 
\beq
\label{eq:105}
\Pi^I(\bq, \Omega) = 2 e^2 v_F \Omega \int_{-\infty}^{\infty} \frac{d \epsilon}{2 \pi i} \frac{d n_F(\epsilon)}{d \epsilon} P_{0 x}(\bq, \epsilon - i \eta, \epsilon + \Omega + i \eta), 
\eeq
and 
\beq
\label{eq:106}
\Pi^{II}(\bq, \Omega) = 4 i  e^2 v_F \int_{-\infty}^{\infty} \frac{d \epsilon}{2 \pi i} n_F(\epsilon) \textrm{Im} P_{0 x}(\bq, \epsilon + i \eta, \epsilon + \Omega + i \eta). 
\eeq 
The 4$\times$4 matrix $P$, whose $0 x$ component we are interested in, is given by 
\beqa
\label{eq:106.1}
P(\bq, - i \eta, \Omega + i \eta)&=&\gamma^{-2} I^{RA}(\bq, \Omega) \cD(\bq, \Omega), \nonumber \\
P(\bq, \epsilon + i \eta, \epsilon + \Omega + i \eta)&=&I^{RR}(\epsilon, \bq, \Omega), 
\eeqa
where 
\beq
\label{eq:106.2}
\cD = (1 - I^{RA})^{-1},
\eeq
is the diffusion propagator and 
\beqa
\label{eq:106.3}
I^{RA}_{\alpha \beta}(\bq, \Omega)&=&\frac{\gamma^2}{2} \tau^{\alpha}_{\sigma_2 \sigma_1} \tau^{\beta}_{\sigma_3 \sigma_4} \int \frac{d^3 k}{(2 \pi)^3} G^R_{\sigma_1 \sigma_3}(\bk + \bq, \Omega) G^A_{\sigma_4 \sigma_2}(\bk , 0), \nonumber \\
I^{RR}_{\alpha \beta}(\epsilon,\bq, \Omega)&=&\frac{1}{2} \tau^{\alpha}_{\sigma_2 \sigma_1} \tau^{\beta}_{\sigma_3 \sigma_4} 
\int \frac{d^3 k}{(2 \pi)^3} G^R_{\sigma_1 \sigma_3}(\bk + \bq,\epsilon + \Omega) G^R_{\sigma_4 \sigma_2}(\bk, \epsilon). 
\eeqa
The physical meaning of the two distinct contributions to the response function $\Pi^{I,II}(\bq, \Omega)$ is clear from Eqs.~\eqref{eq:105}, \eqref{eq:106}.  
$\Pi^{I}(\bq, \Omega)$ describes the non-equilibrium part of the response that happens at the Fermi surface, as clear from the appearance of the derivative of the Fermi-Dirac distribution function in Eq.~\eqref{eq:105}. This response is 
diffusive when $\Omega \tau \ll 1$ and ballistic in the opposite limit, and is generally affected significanty by the impurity scattering.  We will discuss this in more detail below. 
In contrast, $\Pi^{II}(\bq, \Omega)$ is an equilibrium, nondissipative contribution to the overall response, to which all states below 
the Fermi energy contribute~\cite{Streda}, and which is essentially unaffected by the impurity scattering, as will be seen below.

We will start by evaluating the nonequilibrium part of the response function, $\Pi^I(\bq, \Omega)$. 
Computation of the matrix elements $I^{RA}_{\alpha \beta}(\bq, \Omega)$ is easily done in the standard way, assuming $\epsilon_F  \tau \gg 1$. 
One obtains
\beq
\label{eq:107}
\Pi^I(\bq, \Omega) = i e^2 v_F \Omega \tau g(\epsilon_F) [I^{RA}_{00} \cD_{0x} + I^{RA}_{0x} \cD_{xx} + I^{RA}_{0z} \cD_{zx}], 
\eeq
where we have taken into account that $\cD_{yx} = 0$ by symmetry. 
The relevant matrix elements of the diffusion propagator can be found analytically to first order in $\bq$. One obtains
\beq
\label{eq:108}
\Pi^I(\bq, \Omega) = i e^2 v_F \Omega \tau g(\epsilon_F) \frac{I^{RA}_{0x} (1 - I^{RA}_{zz}) + I^{RA}_{0z} I^{RA}_{zx}}{\Gamma (1- I^{RA}_{xx})}, 
\eeq
where
\beq
\label{eq:108.1}
\Gamma(\bq, \Omega) = (1 - I^{RA}_{00})(1 - I^{RA}_{zz}) - I^{RA}_{0z} I^{RA}_{z0}, 
\eeq
is the determinant  of the $0z$ block of the diffusion propagator, which corresponds to the diffusion of the charge density, 
a conserved quantity (this block decouples from the rest of the diffuson when $\bq \rightarrow 0$). 
This means, in particular, that $\Gamma$ must satisfy an exact Ward identity $\Gamma(0,0) = 0$, which is easily checked to be true. 
\begin{figure}[t]
\subfigure[]{
   \label{fig:4a}
  \includegraphics[width=7cm]{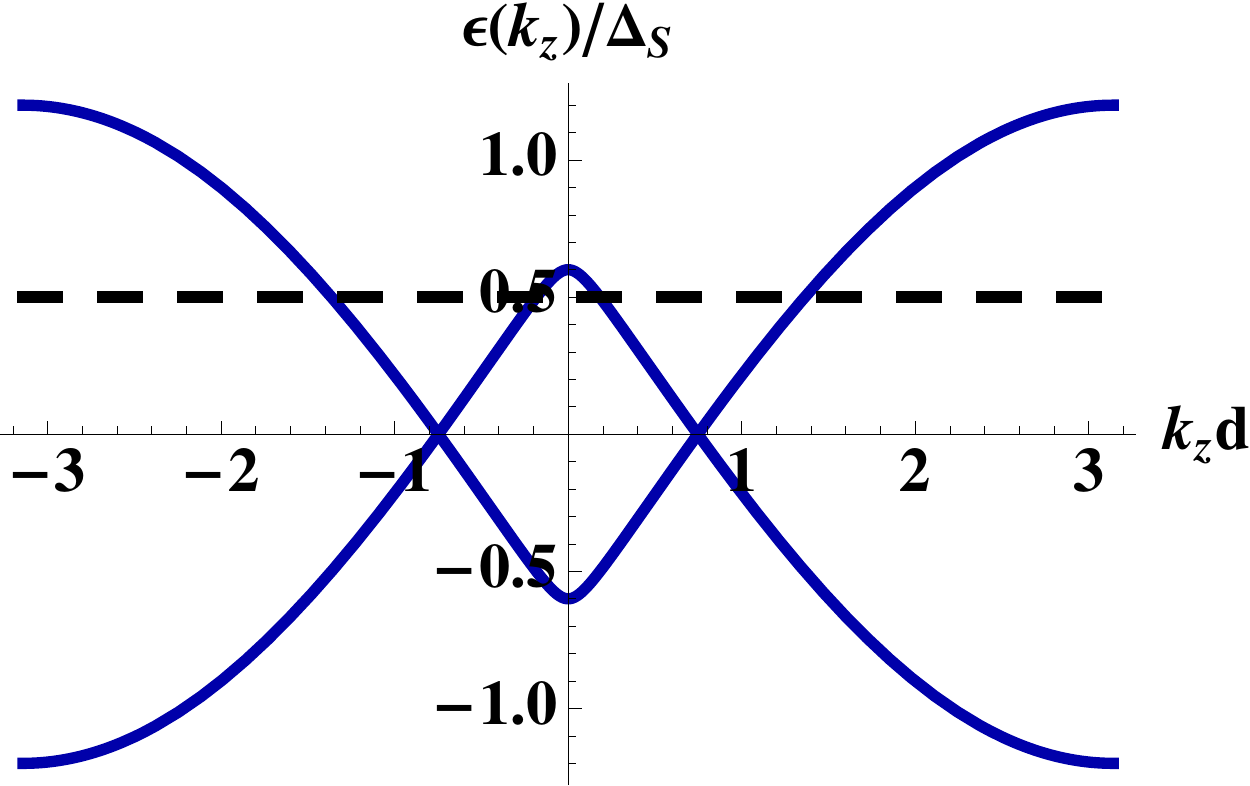}}
\subfigure[]{
  \label{fig:4b}
   \includegraphics[width=7cm]{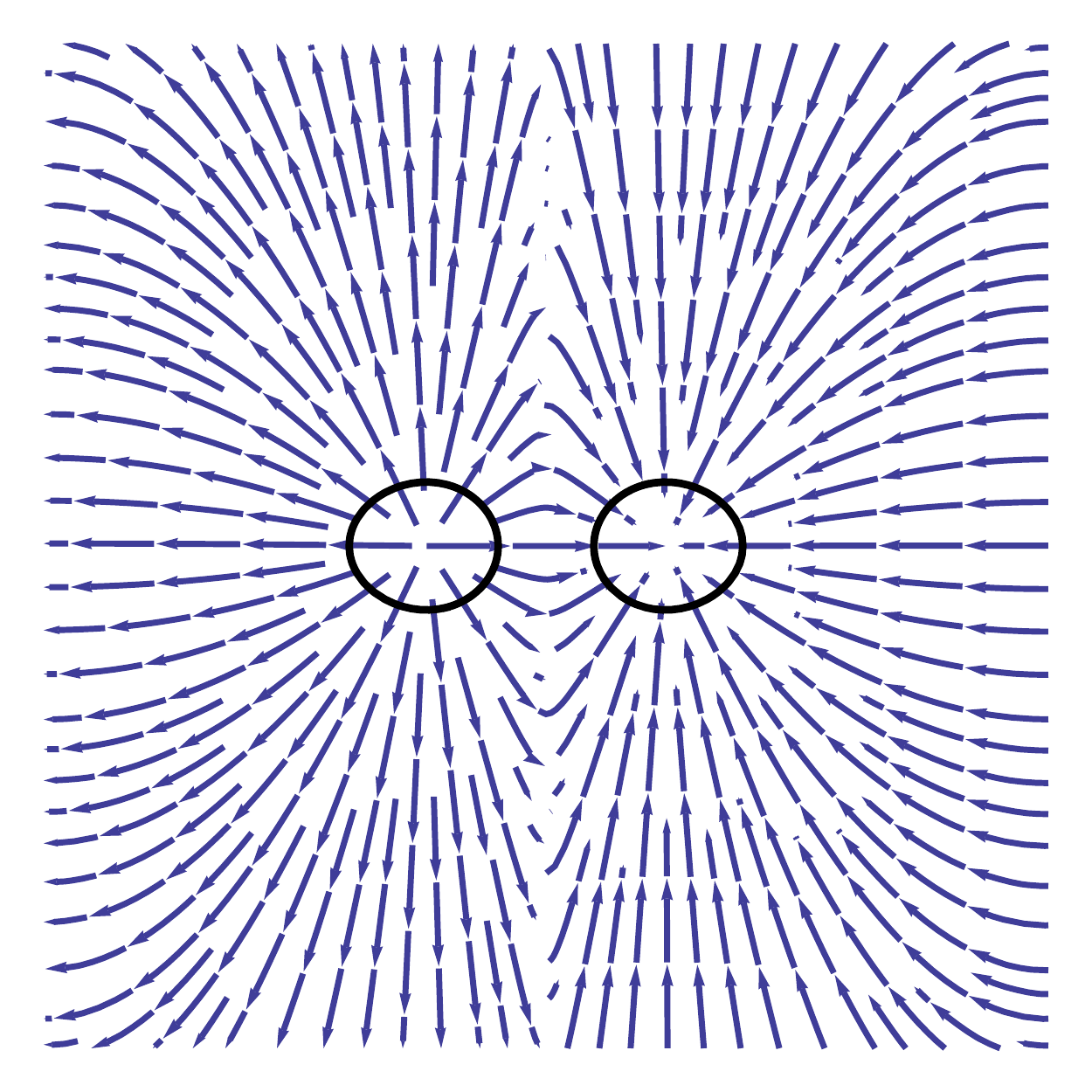}}
  \caption{(a) Plot of the band edges along the $z$-direction in momentum space for the two bands that touch at the Weyl nodes, using $\Delta_D/\Delta_S = -0.9$. 
  (b) Field lines of the Berry curvature 
  in the $k_y = 0$ plane, for the same band structure as in (a). Corresponding Fermi surface section is shown by the two contours, enclosing the Weyl nodes.}
    \label{fig:4}
\end{figure} 
Explicitly, the relevant matrix elements of $I^{RA}(\bq, \Omega)$ to first order in $\bq$ are given by
\beqa
\label{eq:109}
&&I^{RA}_{00} =\left\langle \frac{\tau_+/\tau}{1 - i \Omega \tau_+} \right\rangle, \,\, 
I^{RA}_{0x} = \frac{i v_F q}{2 \epsilon_F} \left\langle \frac{m }{\epsilon_F} \frac{\tau_+/\tau}{1 - i\Omega \tau_+} \right\rangle, \nonumber \\
&&I^{RA}_{0z} = I^{RA}_{z0} = \left\langle \frac{m}{\epsilon_F} \frac{\tau_+/\tau}{1 - i \Omega \tau_+}\right\rangle, \nonumber \\
&&I^{RA}_{zx} = \frac{i v_F q}{4 \epsilon_F} \left\langle \left(1+ \frac{m^2}{\epsilon_F^2}\right) \frac{\tau_+/\tau}{1 - i\Omega \tau_+} \right\rangle, \nonumber \\
&&I^{RA}_{zz} =\left\langle \frac{m^2}{\epsilon_F^2} \frac{\tau_+/\tau}{1 - i \Omega \tau_+} \right\rangle, \nonumber \\
&&I^{RA}_{xx} = \frac{1}{2} \left\langle \left(1- \frac{m^2}{\epsilon_F^2}\right) \frac{\tau_+/\tau}{1 - i\Omega \tau_+} \right\rangle,
\eeqa
where the average over the Fermi surface is defined in the same way as in Eq.~\eqref{eq:103}. 

Expanding $\Gamma(0,\Omega)$ to first order in $\Omega$ and taking the limit $\Omega \rightarrow 0$ at fixed $\tau$, which corresponds to the diffusive 
limit, we obtain
\beq
\label{eq:111}
\Pi_{dif}^I(\bq, 0) =  - \frac{i q \,e^2 v_F^2 g(\epsilon_F)}{2 \epsilon_F}\left\langle \frac{m \tau_+}{\epsilon_F \tau} \right\rangle F[m], 
\eeq
where
\beq
\label{eq:112}
F[m] = \frac{1 + \frac{1}{2} \left\langle \left(1 - \frac{m^2}{\epsilon_F^2} \right) \frac{\tau_+}{\tau}\right\rangle}
{1 - \frac{1}{2} \left\langle \left(1 - \frac{m^2}{\epsilon_F^2} \right) \frac{\tau_+}{\tau}\right\rangle}
\left[\left.\frac{\partial \Gamma(0,\Omega)}{\partial (\Omega \tau)}\right|_{\Omega = 0}\right]^{-1}. 
\eeq 
The explicit form of the functional $F[m]$ is in fact not that important for our purposes, except for the evenness property,
easily seen from Eq.~\eqref{eq:112}: $F[m] = F[-m]$. 
As a consequence, $\Pi_{dif}^I$ is an odd functional of $m$, which will play an important role below.
It is important to note that the charge conservation, whose mathematical consequence is the presence of the diffusion 
pole in $\Pi^I(\bq, \Omega)$, is crucial in obtaining a nonzero result in the diffusive limit in Eq.~\eqref{eq:111}. 
The analogous quantity in the calculation of the spin Hall conductivity, for example, would vanish in the diffusive limit~\cite{Sinova04,Halperin04,Molenkamp04,Molenkamp06}.

It is also of interest to examine the ballistic limit of $\Pi^I$, which corresponds to the case of a clean Weyl metal. In this case 
we send both $\Omega$ and $1/\tau$ to zero in such a way that $\Omega \tau \rightarrow \infty$. 
In this case we obtain
\beqa
\label{eq:113}
\Pi^I_{bal}(\bq,0)&=&\lim_{\Omega \rightarrow 0} \lim_{1/\tau \rightarrow 0} \Pi^I(\bq, \Omega) = 
\lim_{\Omega \rightarrow 0} \lim_{1/\tau \rightarrow 0} i e^2 v_F \Omega \tau g(\epsilon_F) I^{RA}_{0x} \nonumber \\
&=&-\frac{i q \, e^2 v_F^2  g(\epsilon_F)}{2 \epsilon_F} \left\langle \frac{m}{\epsilon_F} \right\rangle, 
\eeqa
which agrees with the clean Weyl metal result Eq.~\eqref{eq:68}. 
A nice feature of the present calculation, compared with the calculation in a clean Weyl metal in Section~\ref{sec:3.2}, is that here 
it is particularly clear that $\Pi^I(\bq, \Omega)$ is associated with states on the Fermi surface. 
As seen from Eqs.~\eqref{eq:111} and~\eqref{eq:113}, the difference between $\Pi^I_{dif}$ and $\Pi^I_{bal}$ is 
only quantitative. In the AHE literature, this difference is said to arise from the 
so-called side-jump processes~\cite{Molenkamp06,Nagaosa06,MacDonald07,Sinova10,Niu11}. 
\begin{figure}[t]
  \includegraphics[width=8cm]{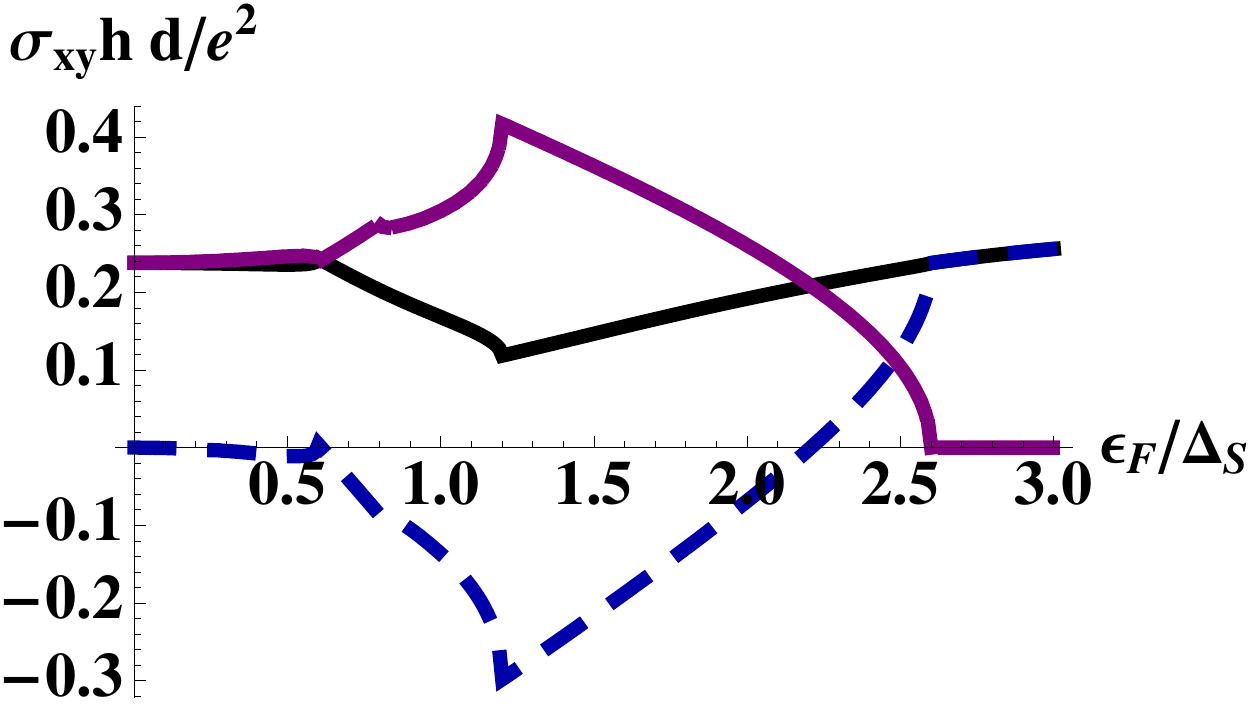}
  \caption{Plot of the total anomalous Hall conductivity (solid line), $\sigma^I_{xy}$ (dashed line) and $\sigma^{II}_{xy}$ (dotted line) versus the Fermi energy for the same system as in Fig.~\ref{fig:4}.
  The plateau-like feature in $\sigma_{xy}$ correlates with the range of the Fermi energies, for which the Fermi surface consists of two separate sheets, each enclosing 
  a single Weyl node.}
  \label{fig:5}
\end{figure}
The final step of the calculation is to evaluate the equilibrium part of the response function, $\Pi^{II}(\bq, \Omega)$. 
In the limit $\epsilon_F \tau \gg 1$ one finds that this part of the response function is unaffected by the impurity scattering and is 
given by
\beqa
\label{eq:114}
\Pi^{II}(\bq, \Omega) = e^2 v_F \int \frac{d^3 k}{(2 \pi)^3} \langle z^{s'}_\bk | z^s_{\bk + \bq} \rangle \langle z^s_{\bk + \bq} | \tau^x | z^{s'}_\bk \rangle \frac{n_F(\xi^s_{\bk + \bq} - \Omega) - n_F(\xi^{s'}_\bk)}{\Omega - \xi^s_{\bk + \bq} + \xi^{s'}_\bk}, \nonumber \\
\eeqa
where summation over the band indices $s, s'$ is again implicit. 
Evaluating Eq.~\eqref{eq:114} in the small $\Omega$ and $q$ limit gives 
\beq
\label{eq:115}
\Pi^{II}(\bq, 0) = \frac{-i q \, e^2}{8 \pi^2} \int_{-\pi/d}^{\pi/d} d k_z \textrm{sign}[m(k_z)] \left\{1 - \Theta[\epsilon_F  - |m(k_z)|]\right\}.
\eeq
The first term in Eq.~\eqref{eq:115} arises from the completely filled bands, while the second is the contribution of the incompletely filled bands. 

We can now finally evaluate the anomalous Hall conductivity. We will focus on the diffusive limit results, as ballistic limit has already been discussed in Section~\ref{sec:3}. 
At this point we will also explicitly include the contribution of both $t = \pm$ pairs of bands, which simply amounts to restoring the index $t$ in $m_t$, and 
summing over $t$. 
Using Eqs.~\eqref{eq:100},\eqref{eq:111} and \eqref{eq:115}, and remembering that $A_0 \rightarrow i A_0$ upon Wick rotation to the real time, we obtain
\beq
\label{eq:116}
\sigma^I_{xy} = \frac{e^2 v_F^2}{2 \epsilon_F} \sum_t g_t(\epsilon_F) \left\langle \frac{m_t \tau_{+ t}}{\epsilon_F \tau_t} \right\rangle F[m_t], 
\eeq
and 
\beq
\label{eq:117}
\sigma^{II}_{xy} = \frac{e^2}{8 \pi^2} \sum_t \int_{-\pi/d}^{\pi/d} d k_z \textrm{sign}[m_t(k_z)] \left\{1 - \Theta[\epsilon_F  - |m_t(k_z)|]\right\}.
\eeq
Since $m_+(k_z)$ is positive throughout the first BZ, while $m_-(k_z)$ changes sign at the Weyl nodes, the first term 
in Eq.~\eqref{eq:117}, which comes from completely filled bands, gives a universal (almost) quantized contribution
\beq
\label{eq:118}
\sigma^{quant}_{xy} = \frac{e^2 {\cal K}}{4 \pi^2}, 
\eeq
where ${\cal K}$ is the distance between the Weyl nodes, which is the same as the clean well metal result, Eq.~\eqref{eq:69}. 
This equation also describes the cases when the Weyl nodes are absent, 
in which case $\sigma^{quant}_{xy}$ is truly quantized since ${\cal K} = 0, G$, where $G= 2\pi /d$ is a reciprocal lattice vector. 

The most important new result of this subsection comes from examining the remaining, non-quantized parts of $\sigma_{xy}$. 
Suppose we have a situation when the Weyl nodes are present and $\epsilon_F$, while not zero, is not too far from it, 
as shown in Fig.~\ref{fig:4}.
Recall that at the location of the Weyl nodes $m_-(k_z) = b - \Delta(k_z)$ changes sign. 
This implies that, as long as ${\cal K} \left.\frac{d \Delta}{d k_z}\right|_{k_z = k_{z0}} \gg \epsilon_F$, 
where $k_{z0}$ is the location of a given Weyl node, the average of any odd function of $m_-(k_z)$ over the Fermi surface will vanish. 
This means that in such a situation, which we call {\em Weyl metal}, all contributions to the anomalous Hall conductivity, 
associated with incompletely filled bands, will vanish and $\sigma_{xy}$ attains a universal value, characteristic 
of Weyl semimetal $\sigma_{xy} = \sigma_{xy}^{quant}$, where $\sigma_{xy}^{quant}$ is given by Eq.~\eqref{eq:118}. 
This may be seen explicitly in Fig.~\ref{fig:5}. 
Note that the linear dispersion sufficiently close to Weyl nodes is a topological property, 
in the sense that it follows directly and exclusively  from the existence of a nonzero topological charge by the so-called Atiyah-Bott-Shapiro construction~\cite{Horava}.  

To understand this result physically, recall that the Weyl nodes are monopole sources of the Berry curvature $\bOmega_\bk$. 
In a clean metal, the anomalous Hall conductivity $\sigma_{xy}$ is given by the integral of the $z$-component of the Berry curvature over all 
occupied states $\sigma_{xy} = e^2 \int \frac{d^3 k}{(2 \pi)^3} n_F(\epsilon_\bk) \Omega^z_\bk$.
However, as clear from Fig.~\ref{fig:5}, when the Fermi surface breaks up into disconnected sheets, enclosing individual nodes, 
the contribution of the states, enclosed by the Fermi surface, to this integral will always be very small, vanishing exactly in the limit when the band dispersion away from the nodes may be taken to be exactly linear.
A useful analogy here is with the electric field of a dipole. A pair of Weyl nodes is like a dipole of two topological charges. Its field has a well-defined and nonzero on average 
$z$-component at large distances from the dipole. At short distances, however, the field is that of individual charges, which winds around the location 
of each charge and thus any particular component of it averages to zero. 

We have thus demonstrated that the AHE in Weyl metals has a purely intrinsic origin and can be associated entirely with the Weyl nodes, just 
as in the case of a Weyl semimetal, when the Fermi energy coincides with the nodes and the Fermi surface is absent~\cite{Burkov14-1}. 
This is in contrast to an ordinary ferromagnetic metal, in which the anomalous Hall conductivity always has both a significant Fermi surface contribution and 
an extrinsic contribution. This property of magnetic Weyl metals may be thought of as being a consequence of emergent chiral symmetry, as discussed in 
Section~\ref{sec:3.2}. 

\subsection{Magnetoresistance in Weyl metals}
\label{sec:4.2}
In this section we will discuss measurable consequences of the Chiral Magnetic Effect, discussed in Section~\ref{sec:3.3}, in the diffusive transport regime. 
As we will demonstrate, CME in this regime leads to a novel type of magnetoresistance: negative and quadratic in the magnetic field, first discovered by Son and Spivak~\cite{Spivak12,Son12} in TR-symmetric Weyl semimetals. As we will show below, this novel magnetoresistance is in fact a universal feature of all types of Weyl semimetals and 
may be regarded as their smoking-gun transport characteristic~\cite{Burkov14-2}. 

We start from the axial charge density operator, given by Eq.~\eqref{eq:74}.
We now ask the following question: does $\hat n_a$ represent a conserved quantity, as it would in a low-energy model of Weyl semimetal? 
Only when $n_a$ is conserved, or nearly conserved, may we expect it to contribute significantly to observable phenomena, at least at long times on 
long length scales. 
To answer this we need to evaluate the commutator of $\hat n_a$ with the Hamiltonian $\cH(\bk)$. 
As before, it is convenient at this point to apply the following canonical transformation to all the operators: $\sigma^{\pm} \rightarrow \tau^z \sigma^{\pm},\,\, \tau^{\pm} \rightarrow
\sigma^z \tau^{\pm}$. Evaluating the commutator at the Weyl node locations, we now obtain
\beq
\label{eq:119}
\left[\cH(\bk), \hat n_a\right]_{k^z_{\pm}} =  i \frac{b^2 - \Delta_D^2 + \Delta_S^2}{\Delta_S}  \tau^z \sigma^z. 
\eeq
This means that $n_a$ may indeed be a conserved quantity in the Weyl semimetal or weakly-doped Weyl metal, provided $\Delta_D \geq \Delta_S$ and $b = \sqrt{b_{c1} b_{c2}}$, 
i.e. the magnitude of the spin splitting is exactly the geometric mean of its lower- and upper-critical values, at which the transitions out of the Weyl semimetal phase occur. 
Otherwise, the commutator is nonzero and $n_a$ is not conserved. However, as will be shown below, the relevant relaxation time may in fact still be long, even when the above 
condition is not exactly satisfied, in which case the axial charge density is still a physically meaningful quantity. 
The near-conservation of the axial charge density $n_a$ may be viewed as a consequence of an emergent low-energy {\em chiral symmetry}, which is an important 
characteristic feature of Weyl semimetals. Both the quadratic negative magnetoresistance, and the fully intrinsic disorder-independent AHE, discussed above, are consequences 
of this emergent symmetry.   

We now want to derive hydrodynamic transport equations for both the axial charge density $n_a(\br, t)$ and the total charge density $n(\br, t)$. 
As will be shown below, what is known as chiral anomaly will be manifest at the level of these hydrodynamic equations as a coupling between 
$n_a$ and $n$ in the presence of an external magnetic field. This coupling leads to significant observable magnetotransport effects, provided the 
axial charge relaxation time, calculated below, is long enough. As we will show explicitly below, long axial charge relaxation time is a direct 
consequence of the emergent chiral symmetry, which exists in a Weyl metal at low energies. 

To proceed with the derivation, we add a constant uniform magnetic field in the $\hat z$ direction $\bB = B \hat z$ and a scalar impurity potential $V(\br)$, 
whose precise form will be specified later. 
Adopting Landau gauge for the vector potential $\bA = x B \hat y$, the second-quantized Hamiltonian of our system may be written as
\beqa
\label{eq:120}
H&=&\sum_{n a k_y k_z} \epsilon_{n a}(k_z) c^\dg_{n a k_y k_z} c^\pdg_{n a k_y k_z}  \nonumber \\
&+&\sum_{n a k_y k_z, n'  a'  k_y' k_z'} \langle n, a, k_y, k_z | V | n', a', k_y', k_z' \rangle c^\dg_{n a k_y k_z} c^\pdg_{n' a' k_y' k_z'}. 
\eeqa
Here $\epsilon_{n a}(k_z)$ are Landau-level (LL) eigenstate energies of a clean multilayer in magnetic field, $n=0,1,2,\ldots$ is the main LL
index, $k_y$ is the Landau-gauge intra-LL orbital quantum number, $k_z$ is the conserved component of the crystal momentum 
along the $z$-direction, and $a = (s, t)$ is a composite index (introduced for 
compactness of notation), consisting of $s = \pm$, which labels the electron- ($s = +$) and hole- ($s = -$) like sets of Landau levels, and $t = \pm$, which 
labels the two components of a Kramers doublet of LLs, degenerate at $b = 0$.
The LL eigenstate energies and the corresponding eigenvectors have been derived in Section~\ref{sec:3.3}. 

To proceed, we will make the standard assumption, which we also made in the previous subsection, that the impurity potential obeys Gaussian distribution, with $\langle V(\br) V(\br') \rangle = \gamma^2 \delta(\br - \br ')$.
To simplify calculations further we will also assume that the momentum transfer due to the impurity scattering is smaller than the size of the BZ, i.e. $| k_z - k_z' | d \ll 1$. 
In this case $\langle u^t(k_z)| u^{t'}(k_z') \rangle \approx \delta_{t t'}$, i.e. the $t$ quantum number may be assumed to be approximately preserved during the impurity scattering. 
\begin{figure}[t]
\begin{center}
  \includegraphics[width=12cm]{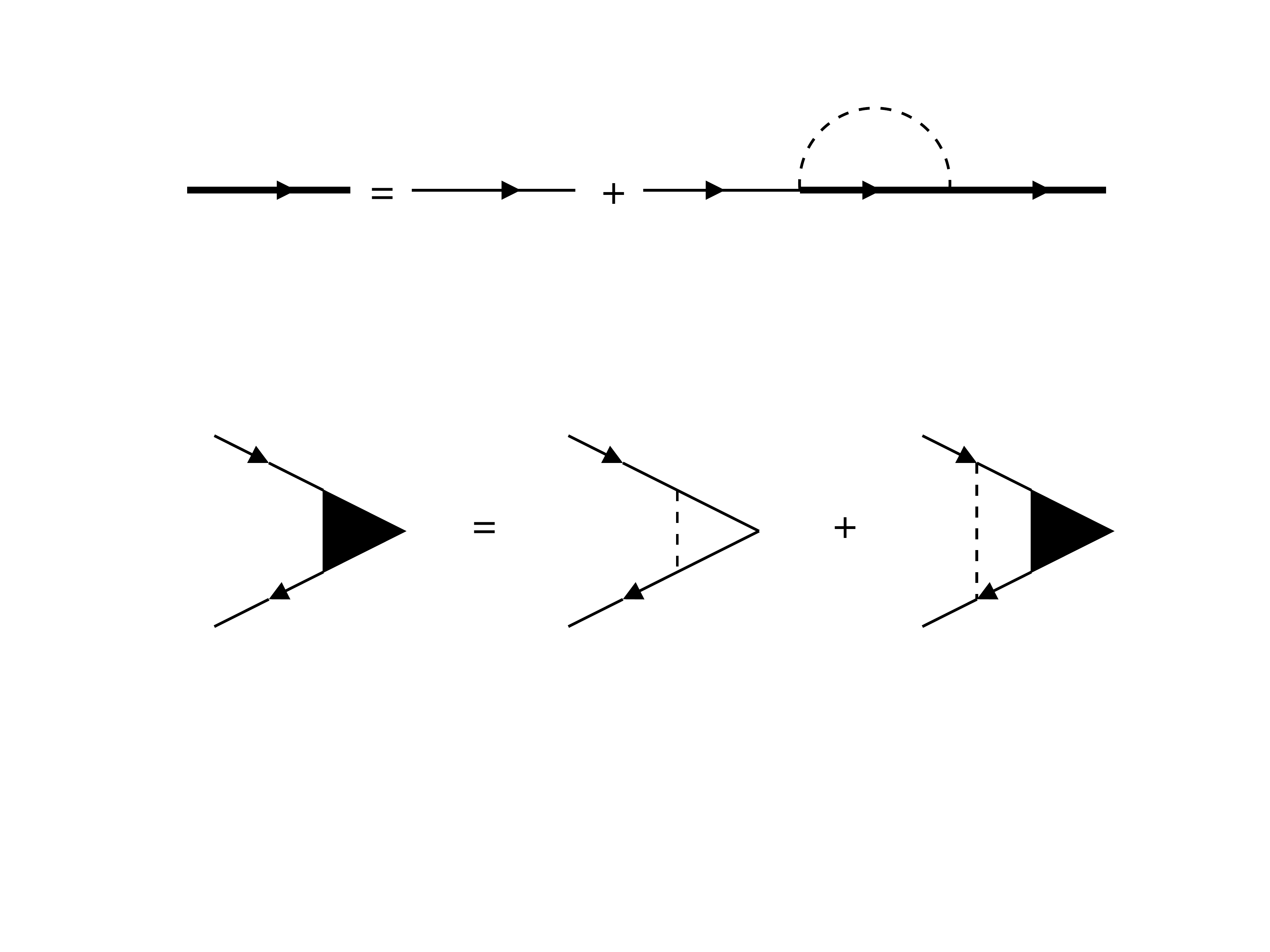}
\end{center}
\vspace{-2cm}
  \caption{Graphical representation of the impurity-averaged Green's function within SCBA and of the diffusion propagator.}  
  \label{fig:3.6}
\end{figure} 
We treat the impurity scattering in the standard self-consistent Born approximation (SCBA), illustrated in Fig.~\ref{fig:3.6}. 
The retarded SCBA self-energy satisfies the equation
\beq
\label{eq:122}
\Sigma^R_{n a k_y k_z }(\omega) = \frac{1}{L_z} \sum_{n' a' k_y'  k_z'} \langle |\langle n, a, k_y, k_z| V |n', a', k_y', k_z'\rangle|^2 \rangle G^R_{n' a' k_y' k_z'}(\omega), 
\eeq
We will assume that the Fermi energy $\epsilon_F$ is positive, i.e. the Weyl semimetal is electron-doped, and large enough that the impurity-scattering-induced broadening of the density of states is small on the scale of the Fermi energy $\epsilon_F$~\cite{Biswas14}. We can then restrict ourselves to the electron-like states with $s=+$ (we will drop the $s$ index henceforth for brevity), and easily solve the SCBA equation analytically. We obtain
\beq
\label{eq:123}
\textrm{Im} \Sigma^R_{n t k_z} \equiv -\frac{1}{2 \tau_t(k_z)} = - \frac{1}{2 \tau} \left[1 + \frac{m_t(k_z) \langle m_t \rangle}{\epsilon_F^2} \right], 
\eeq
where $1/ \tau = \pi \gamma^2 g(\epsilon_F)$ and 
\beq
\label{eq:124}
g(\epsilon_F) = \frac{1}{2 \pi \ell_B^2} \int_{-\pi/d}^{\pi/d} \frac{d k_z}{2 \pi} \sum_{n t} \delta[\epsilon_{n t}(k_z) - \epsilon_F],
\eeq
is the density of states at Fermi energy. 
We have also introduced the Fermi-surface average of $m_t(k_z)$ as
\beq
\label{eq:125}
\langle m_t \rangle = \frac{1}{2 \pi \ell_B^2 g(\epsilon_F)} \int_{-\pi/d}^{\pi/d} \frac{d k_z}{2 \pi} \sum_{n t} m_t(k_z) \delta[\epsilon_{n t}(k_z) - \epsilon_F]. 
\eeq

We are interested in hydrodynamic, i.e. long-wavelength, low-frequency density response of our system. 
As is well-known~\cite{Altland}, the relevant information is contained in the diffusion propagator, or diffuson $\cD$, given 
by the sum of ladder impurity-averaging diagrams.
This is evaluated in the standard manner and we obtain
\beq
\label{eq:126}
\cD^{-1} (\bq, \Omega)= 1 - I(\bq, \Omega), 
\eeq
where $I$ is a $16 \times 16$ matrix, given by
\beqa
\label{eq:127}
I_{\alpha_1 \alpha_2, \alpha_3 \alpha_4}(\bq, \Omega) = \frac{\gamma^2}{L_x L_y L_z} \int d^3 r d^3 r' e^{-i \bq \cdot (\br - \br')} G^R_{\alpha_1 \alpha_3}(\br, \br'| \Omega) G^A_{\alpha_4 \alpha_2}(\br', \br | 0), \nonumber \\
\eeqa
where we have introduced a composite index $\alpha = (\sigma, \tau)$ to simplify the notation. 
The impurity-averaged Green's functions $G^{R,A}$ are given by
\beq
\label{eq:128}
G^{R,A}_{\alpha \alpha'}(\br, \br' |\Omega) = \sum_{n t k_y k_z} \frac{\langle \br, \alpha | n,t,k_y,k_z \rangle \langle n,t,k_y,k_z | \br', \alpha' \rangle}
{\Omega - \xi_{n t}(k_z) \pm i / 2 \tau_t(k_z)}, 
\eeq
where $\xi_{n t}(k_z) = \epsilon_{n t}(k_z) - \epsilon_F$. 

In general, the evaluation of Eq.~\eqref{eq:127} is a rather complicated task, primarily due to the fact that the impurity scattering will mix different LLs. 
At this point we will thus specialize to the case of transport along the $z$-direction only, as this is where we can expect 
chiral anomaly to be manifest. In this case the contributions of different LLs to Eq.~\eqref{eq:127} decouple. 
Setting $\bq = q \hat z$, we obtain
\beqa
\label{eq:129}
&&I_{\alpha_1 \alpha_2, \alpha_3 \alpha_4} (q, \Omega)  = \frac{\gamma^2}{2 \pi \ell_B^2 L_z} \sum_{n t t' k_z} \frac{\langle \alpha_1| z^{t}_{n}(k_z + q/2) \rangle \langle z^{t}_{n}(k_z + q/2) | \alpha_3 \rangle}
{\Omega - \xi_{n t}(k_z + q/2) + i/ 2 \tau_t(k_z + q/2)} \nonumber \\
&\times&\frac{\langle \alpha_4 | z^{t'}_{n}(k_z - q/2) \rangle \langle z^{t'}_{n}(k_z - q/2) | \alpha_2 \rangle}{-\xi_{n t'}(k_z - q/2) - i/ 2 \tau_t'(k_z - q/2)}, 
\eeqa
which is much easier to evaluate.

As mentioned above, $I$ and $\cD^{-1}$ are large $16 \times 16$ matrices, which contain a lot of information of no interest to us. 
We are interested only in hydrodynamic physical quantities, with long relaxation times. All such quantities need to be identified, 
if they are expected to be coupled to each other. 
One such quantity is obviously the total charge density $n(\br, t)$, which has an infinite relaxation time due to the exact conservation of particle number. 
Another is the axial charge density $n_a(\br, t)$, which, as discussed above, may be almost conserved under certain conditions. 
On physical grounds, we expect no other hydrodynamic quantities to be present in our case. We are thus only interested in the $2 \times 2$ block of the matrix $\cD^{-1}$, which corresponds to the coupled evolution of the total and the axial charge densities. 
To separate out this block, we apply the following transformation to the inverse diffuson matrix
\beq
\label{eq:130}
\cD^{-1}_{a_1 b_1,  a_2 b_2} = \frac{1}{2} (\sigma^{a_1} \tau^{b_1})_{\alpha_2 \alpha_1} \cD^{-1}_{\alpha_1 \alpha_2, \alpha_3 \alpha_4}
(\sigma^{a_2} \tau^{b_2})_{\alpha_3 \alpha_4}, 
\eeq
where $a_{1,2}, b_{1,2} = 0, x, y, z$. The components of interest to us are $a_{1,2} = b_{1,2} = 0$ which corresponds to the total charge density, $a_{1,2} = 0, b_{1,2} = y$, 
which corresponds to the axial charge density, and the corresponding cross-terms.   

We will be interested, as mentioned above, in the hydrodynamic regime, which corresponds to low frequencies and long wavelengths, i.e. $\Omega \tau \ll 1$ and 
$v_F q \tau \ll 1$. We will also assume that the magnetic field is weak, so that $\omega_B \ll \epsilon_F$. 
Finally, we will assume that the Fermi energy is close enough to the Weyl nodes, so that only the $t = -$ states participate in transport and 
$\langle m_- \rangle \approx 0$, since $m_-(k_z)$ changes sign at the nodes~\cite{Burkov14}. 

In accordance with the above assumptions, we expand the inverse diffusion propagator to leading order in $\Omega \tau$, $v_F q \tau$ and $\omega_B/ \epsilon_F$ 
and obtain after a straightforward but lengthy calculation 
\beqa
\label{eq:131}
\cD^{-1}(q, \Omega) = \left(
\begin{array}{cc}
 -i \Omega \tau + D q^2 \tau & -i q \Gamma \tau \\
 - i q \Gamma \tau & -i \Omega \tau + D q^2 \tau + \tau/ \tau_a
 \end{array}
 \right). \nonumber \\
 \eeqa 
Here $D = \tilde v_F^2 \tau \langle m_-^2 \rangle/ \epsilon_F^2$ is the charge diffusion constant, associated with the diffusion in the $z$-direction, 
$\Gamma = e B / 2 \pi^2 g(\epsilon_F)$ is the total charge-axial charge coupling coefficient and 
\beq
\label{eq:132}
\frac{1}{\tau_a} = \frac{1 - (\tilde v_F/ \Delta_S d)^2}{(\tilde v_F/ \Delta_S d)^2 \tau}, 
\eeq
is the axial charge relaxation rate. 
Several comments are in order here. First, an important thing to note is that the off-diagonal matrix elements in $\cD^{-1}$, proportional 
to $B$ and responsible for the total charge to axial charge coupling, come entirely from the contribution of the $n = 0$ LL. 
The remaining matrix elements arise from the contribution of all the $n \geq 1$ Landau levels and we have taken the limit $B \rightarrow 0$ 
after summing over the LLs, i.e. left only the leading term in the $\omega_B/ \epsilon_F$ expansion. The next-to-leading term results in 
a negative correction to the diffusion coefficient, proportional to $B^2$, which corresponds to the well-known classical positive magnetoresistance. 
We have ignored this correction here, but will comment on its effects later. 
Second, we note that the axial charge relaxation rate $1/\tau_a \geq 0$, as it should be, and vanishes when $\tilde v_F = \Delta_S d$. 
It is easy to see that this is identical to the condition of the vanishing of the commutator of the axial charge operator with the Hamiltonian Eq.~\eqref{eq:119}, 
again as it should be.
Henceforth we will assume that this condition is nearly satisfied so that $\tau_a \gg \tau$. 
Finally, the situation when $\tilde v_F = \Delta_S d$ and thus $1/\tau_a$ appears to vanish, actually needs to be treated with some care. Namely, 
the condition $\tilde v_F = \Delta_S d$ may be satisfied exactly only in the limit $\epsilon_F \rightarrow 0$. The Fermi velocity  depends on the Fermi energy as~\cite{Pesin14}
\beq 
\label{eq:132.1}
\tilde v_F(\epsilon_F) = \frac{d}{2 (b + \epsilon_F)} \sqrt{[(b + \epsilon_F)^2 - b_{c1}^2] [b_{c2}^2 - (b + \epsilon_F)^2]}. 
\eeq
When $b = \sqrt{b_{c1} b_{c2}}$ and thus $\tilde v_F(0) = \Delta_S d$, the Fermi energy dependence of $\tilde v_F$ needs to be taken into account. 
Expanding to leading non vanishing order in $\epsilon_F$ we obtain in this case
\beq
\label{eq:132.2}
\frac{1}{\tau_a} = \frac{\epsilon_F^2}{\Delta_S^2 \tau}, 
\eeq
i.e. $1/\tau_a$ is in fact always finite, but may be very small. We can estimate the minimal value of the axial charge relaxation rate by 
setting $\epsilon_F \approx 1/\tau$ in Eq. ~\eqref{eq:132.2}, which gives $(\tau/\tau_a)_{min} \approx 1/ (\Delta_S \tau)^2$. 

We may now write down the coupled diffusion equations for the total and axial charge densities, which correspond to the propagator Eq.~\eqref{eq:131}. 
These equations read
\beqa
\label{eq:133}
\frac{\partial n}{\partial t}&=&D \frac{\partial^2 n}{\partial z^2} + \Gamma \frac{\partial n_a}{\partial z}, \nonumber \\
\frac{\partial n_a}{\partial t}&=&D \frac{\partial^2 n_a}{\partial z^2} - \frac{n_a}{\tau_a} + \Gamma \frac{\partial n}{\partial z}. 
\eeqa
Manifestation of chiral anomaly in these equations is the coupling between the total and the axial charge densities, proportional to the 
applied magnetic field. 
Since the total particle number is conserved, the right-hand side of the first of Eqs.~\eqref{eq:133} must be equal to minus the divergence of the total particle current. 
Then we obtain the following expression for the density of the charge current in the $z$-direction
\beq
\label{eq:134}
j = - \frac{\sigma_0}{e} \frac{\partial \mu}{\partial z} - \frac{e^2 B}{2 \pi^2} \mu_a, 
\eeq
where $\sigma_0 = e^2 g(\epsilon_F) D$ is the zero-field diagonal charge conductivity, $\mu$ and $\mu_a$ are the total and axial electrochemical potentials 
and we have used $\delta n = g(\epsilon_F) \delta \mu$, $\delta n_a = g(\epsilon_F) \delta \mu_a$. The last relation is valid when $\tilde v_F/ \Delta_S d$ is close
to unity, as seen from Eq.~\eqref{eq:75}. 
Thus chiral anomaly manifests in an extra contribution to the charge current density, proportional to the magnetic field and the axial electrochemical potential. 
This is known as chiral magnetic effect (CME) in the literature~\cite{Kharzeev}. 
It is important to realize that the second term in Eq.~\eqref{eq:134} does not by any means imply that an equilibrium current may be driven by an applied 
magnetic field, despite appearances. The axial chemical potential $\mu_a$, appearing in Eq.~\eqref{eq:134}, is a purely nonequilibrium quantity. 
If an equilibrium energy difference, $\mu_{a0}$,  exists between the Weyl nodes due to explicitly broken inversion symmetry~\cite{Zyuzin12-2}, then it is the difference $\mu_a - \mu_{a0}$ 
that enters in Eq.~\eqref{eq:134}. We have explicitly considered an inversion-symmetric Weyl metal here, in which case $\mu_{a0} = 0$. 

To find measurable consequences of the CME contribution to the charge current, we consider a steady-state situation, with a fixed current density $j$ flowing through the 
sample in the $z$-direction. We want to find the corresponding electrochemical potential drop and thus the conductivity. 
Assuming the current density is uniform, we obtain from the second of Eqs.~\eqref{eq:133}
\beq
\label{eq:135}
n_a = \Gamma \tau_a \frac{\partial n}{\partial z}, 
\eeq 
which is the nonequilibrium axial charge density, induced by the current and the corresponding electrochemical potential gradient. 
Substituting this into the expression for the charge current density Eq.~\eqref{eq:134}, we finally obtain the following result
for the conductivity
\beq
\label{eq:136}
\sigma = \sigma_0 + \frac{e^4 B^2 \tau_a}{4 \pi^4 g(\epsilon_F)}. 
\eeq
In the limit when $\epsilon_F$ is not far from the Weyl nodes, such that the dispersion may be assumed to be linear, 
we have $g(\epsilon_F) = \epsilon_F^2/ \pi^2 v_F^2 \tilde v_F$, which gives
\beq
\label{eq:137}
\Delta \sigma = \sigma - \sigma_0 = \frac{e^2 \tilde v_F \tau_a}{(2 \pi v_F)^2} \left(\frac{e^2 v_F^2 B}{\epsilon_F} \right)^2,
\eeq
which agrees with the Son and Spivak result~\cite{Spivak12,Son12}. 
Thus we see that a measurable consequence of CME is a positive magnetoconductivity, proportional to $B^2$ in the 
limit of a weak magnetic field. 
This of course needs to be compared with the classical negative magnetoconductivity, which is always 
present and arises from the $B^2$ corrections to the diffusion constant $D$, which we have neglected
\beq
\label{eq:138}
\frac{\Delta \sigma_{c \ell}}{\sigma_0} \sim - (\omega_c \tau)^2, 
\eeq 
where $\omega_c = e v_F^2 B/ \epsilon_F$ is the cyclotron frequency. 
This gives 
\beq
\label{eq:139}
\left|\frac{\Delta \sigma}{\Delta \sigma_{c \ell}}\right| \sim \frac{\tau_a/\tau}{(\epsilon_F \tau)^2}. 
\eeq
Thus the CME-related positive magnetoconductivity will dominate the classical negative magnetoconductivity, provided 
$\tau_a$ is long enough. 

We have so far ignored the Zeeman effect due to the applied magnetic field. 
Its effect is to modify the spin-splitting parameter $b$ as $b \rightarrow b + g \mu_B B/2$. 
In principle, the dependence on $b$ does enter into our final results through the dependence 
of the Fermi velocity $\tilde v_F$ on $b$. Naively, this will then generate an additional {\em linear} magnetoconductivity, which 
may be expected to dominate the quadratic one at small fields.
However, the condition of large $\tau_a$, which is the same as $\tilde v_F/ \Delta_S d \approx 1$, is equivalent to 
the condition $b_{c1} \ll b \ll b_{c2}$, in which case the dependence of $\tilde v_F$ on $b$ becomes negligible. 
Thus, in the regime in which the positive magnetoconductivity dominates the negative classical one, and is thus observable, 
one may also expect a negligible linear magnetoconductivity in any type of Weyl metal. 
\section{Conclusions}
\label{sec:5}
In this paper we have provided an overview of transport phenomena in Weyl metals, which may be attributed to chiral anomaly, 
in particular the semi-quantized intrinsic Anomalous Hall Effect and the Chiral Magnetic Effect, which manifests in negative quadratic in the 
magnetic field longitudinal magnetoresistance. 

The AHE is generally present in any ferromagnetic metal. 
In a generic FM metal this is a complicated phenomenon, with many physically distinct sources contributing to the 
final observed effect. In particular, {\em extrinsic} contribution to AHE, arising from impurity scattering, is always at least 
of the same order of magnitude, and often significantly larger, than the more theoretically appealing and universal {\em intrinsic} one. 
In a Weyl metal, however, as we have demonstrated in this paper, AHE has a purely intrinsic origin. As long as the Fermi energy is 
not too far from the Weyl nodes, such that the band dispersion may be taken to be linear and the chiral symmetry is present to a good approximation, 
the anomalous Hall conductivity of a Weyl metal 
turns out to be given exactly by the semi-quantized expression of a pure undoped Weyl semimetal, where it simply measures separation 
between the Weyl nodes in momentum space in units of $e^2/h$. 
Both the extrinsic, and even the part of the intrinsic contribution, associated with the Fermi surface (i.e. incompletely filled bands), vanish identically. 
We have connected this property with the geometry of the Berry curvature field in the vicinity of the Weyl nodes, which in turn is a direct consequence
of their topology, through the Atiyah-Bott-Shapiro construction. 
FM Weyl metal is thus distinguished from other FM metals by the fact that its AHE is of purely intrinsic origin. 

Perhaps even more important in the context of Weyl metals is the Chiral Magnetic Effect and the associated negative quadratic 
magnetoresistance. This may be expected to occur in any type of Weyl metal, either characterized by broken time reversal symmetry
or broken inversion (or both). It is thus of particular importance for the experimental characterization of Weyl metals. 
The negative quadratic longitudinal magnetoresistance should in fact be observed even in Dirac semimetals, which have already 
been realized experimentally. The theory, presented in Section~\ref{sec:4.2} is directly applicable to this case, if $\Delta_S = \Delta_D$ 
limit is taken and the spin-splitting $b$ is identified with the Zeeman splitting $b = g \mu_B B /2$. 
In fact, this effect appears to have been observed very recently in a 3D Dirac semimetal material ZrTe$_5$~\cite{Kharzeev14}. 

What has been presented in this paper is by no means a complete story of the possible observable manifestations of chiral anomaly 
in Weyl metals. Other related phenomena, such as linear high-field magnetoresistance~\cite{Aji12}, plasmon-magnon coupling~\cite{Liu13},
nonlocal transport~\cite{Pesin13}, anomalous density response~\cite{Pesin14}, anomalous thermoelectric response~\cite{Fiete14}, and others~\cite{Parameswaran12,Grushin12,Balatsky13,Goswami13,Qi13,Xiao14,Zyuzin14,Chang14,Abanin14,Miransky14,Nomura14,Hughes14}, have been discussed in the literature. 
One of the possible promising directions for future research in this area is the interplay of chiral anomaly and superconductivity in Weyl metals~\cite{Balents12,Moore12,Aji14}. 

\section{Acknowledgments}
We would like to thank L. Balents, Y. Chen, M.D. Hook, I. Panfilov, D.A. Pesin, S. Wu, and A.A. Zyuzin for collaboration on the topics, covered 
in this review, and closely related ones. We also thank Y. Ando and B.Z. Spivak for useful discussions. Financial support was provided by NSERC of Canada.

\end{document}